\pdfoutput=1
\documentclass[10pt]{article}
\usepackage{amsmath,amssymb,amsthm,amsfonts}
\usepackage{graphicx}
\usepackage{xcolor}
\usepackage[authoryear,round,semicolon]{natbib}
\numberwithin{equation}{section}
\usepackage{enumitem} 

\usepackage[hmargin=1in,vmargin=1in]{geometry}
\usepackage{comment}
\usepackage{subcaption}
\usepackage{booktabs}

{\begin{Sbox}\begin{minipage}}%
{\end{minipage}\end{Sbox}\fbox{\TheSbox}}

\newcommand{\captionfonts}{\small}

\makeatletter  
\long\def\@makecaption#1#2{%
  \vskip\abovecaptionskip
  \sbox\@tempboxa{{\captionfonts #1: #2}}%
  \ifdim \wd\@tempboxa >\hsize
    {\captionfonts #1: #2\par}
  \else
    \hbox to\hsize{\hfil\box\@tempboxa\hfil}%
  \fi
  \vskip\belowcaptionskip}
\makeatother   
\newenvironment{Remark}{
\vspace{10pt}
\noindent{\bf Remark.}\
}{
\leavevmode\unskip\penalty9999 \hbox{}\nobreak\hfill
  \quad\hbox{$\qed$}
 \par \vspace{10pt}
}

\graphicspath{{./figures/}}

\newcommand{\calB}{\mathcal{B}}%
\newcommand{\calD}{\mathcal{D}}%
\newcommand{\calH}{\mathcal{H}}%
\newcommand{\calL}{\mathcal{L}}%
\newcommand{\calS}{\mathcal{S}}%
\newcommand{\calV}{\mathcal{V}}%
\newcommand{\calW}{\mathcal{W}}%
\newcommand{\bfA}{{\bf A}}%
\newcommand{\bfb}{{\bf b}}\newcommand{\bfB}{{\bf B}}%
\newcommand{\bfC}{{\bf C}}%
\newcommand{\bfF}{{\bf F}}%
\newcommand{\bfn}{{\bf n}}%
\newcommand{\bfR}{{\bf R}}%
\newcommand{\bft}{{\bf t}}\newcommand{\bfT}{{\bf T}}%
\newcommand{\bfU}{{\bf U}}%
\newcommand{\bfv}{{\bf v}}\newcommand{\bfV}{{\bf V}}%
\newcommand{\bfx}{{\bf x}}\newcommand{\bfX}{{\bf X}}%
\newcommand{\bfxi}{\boldsymbol{\xi}}%
\newcommand{\bfchi}{\boldsymbol{\chi}}%
\newfont{\tenbfit}{cmmib10}%
\newfont{\svnbfit}{cmmib8}%
\newfont{\tenbfsl}{cmbxti10}
\newcommand{\bfLambda}{\boldsymbol{\Lambda}}%
\newcommand{\bfOmega}{\boldsymbol{\Omega}}
\newfont{\mmit}{cmmi10}
\newfont{\smit}{cmmi9}
\newfont{\bfMit}{cmmi5}
\newfont{\tenbbb}{msbm10}%
\newfont{\svnbbb}{msbm8}%
\newfont{\tenssit}{cmssqi8 at 10pt}%
\newfont{\svnssit}{cmssqi8 at 7pt}%
\newfont{\gothic}{eufm10}%
\newfont{\sgothic}{eufm7}%
\newcommand{\Det}{\hbox{\rm det}\mskip2mu}

\newcommand{\skw}{\hbox{\rm skw}\mskip3mu}
\newcommand{\sym}{\hbox{\rm sym}\mskip3mu}
\newcommand{\Tr}{\hbox{\rm tr}\mskip2mu}

\newcommand{\pards}[2]{\mbox{$\dfrac{\partial #1}{\partial {#2 }}$}}

\newcommand{\vs}{\mskip2mu}
\newcommand{\vvs}{\mskip1mu}
\newcommand{\trans}{{\scriptscriptstyle\mskip-1mu\top\mskip-2mu}}
\newcommand{\inv}{{\scriptscriptstyle\mskip-1mu{-1}\mskip-2mu}}
\newcommand{\invtrans}{{\scriptscriptstyle\mskip-1mu{-\top}\mskip-2mu}}

\newcommand{\Blj}{\mbox{$\Big[\kern-0.275em\Big[$}}
\newcommand{\Brj}{\mbox{$\Big]\kern-0.275em\Big]$}}
\newcommand{\zed}{{\bf 0}}
\newcommand{\id}{{\bf 1}}
\newcommand{\onehalf}{\textstyle{\frac{1}{2}}}
\newcommand{\half}{\textstyle{\dfrac{1}{2}}}

\newcommand{\Div}{\hbox{\rm Div}\mskip2mu}                                      

\newcommand{\B}{\text{B}}
\newcommand{\p}{\text{P}}                                                                                    

\newcommand{\X}{\bfX}
\newcommand{\x}{\bfx}
\newcommand{\y}{\bfchi}
\newcommand{\F}{\bfF}

\newcommand{\C}{\bfC}
\newcommand{\FT}{\bfF^{\trans}}                                                                         

\newcommand{\A}{\bfA}

\newcommand{\Fbar}{\bar\bfF}
\newcommand{\Cbar}{\bar\bfC}
\newcommand{\Bbar}{\bar\bfB}

\newcommand{\tendot}{\mskip-3mu:\mskip-2mu}

\newcommand{\Def}{\overset{\text{def}}{=}}

\newcommand{\thet}{\vartheta}%

\newcommand{\mat}{\text{\tiny R}}%
\newcommand{\T}{\bfT}

\newcommand{\kk}{\text{\tiny K}}%

\newcommand{\Tmat}{\bfT_{\mat}}

\newcommand{\lambdabtilde}{\tilde{\lambda}_b}
\newcommand{\lambdabdot}{\dot{\lambda}_b}
\newcommand{\lambdabar}{\bar{\lambda}}
\newcommand{\lambdab}{\lambda_b}
\newcommand{\prt}{\text{P}}
\newcommand{\symz}{\text{sym}_0}

\newcommand{\iso}{\text{i}}
\newcommand{\vol}{\text{v}}

\newcommand{\dee}{\mbox{$d$}}

\newcommand{\xien}{\mbox{$\bfxi_{\text{en}}$}}

\newcommand{\xidiss}{\mbox{$\bfxi_{\text{diss}}$}}
\newcommand{\fen}{\mbox{$f_\text{en}$}}
\newcommand{\fdiss}{f_\text{diss}}

\newcommand{\varpien} {\mbox{$\varpi_{\text{en}}$}}
\newcommand{\varpidiss} {\mbox{$\varpi_{\text{diss}}$}}
\newcommand{\en}{{\text{en}}}
\newcommand{\diss}{{\text{diss}}}
\newcommand{\deedot}{\dot{d}}

\newcommand{\deetilde}{\tilde{d}}

\newcommand{\Fdot}{\dot\bfF}
\newcommand{\Ftilde}{\tilde\bfF}
\newcommand{\Cdot}{\dot\bfC}

      \newcommand{\gee}{{\text{G}}}

\begin{document}


\title{Progressive damage and rupture in polymers}

\author{
Brandon~Talamini, Yunwei~Mao, Lallit~Anand\thanks{Tel.: +1-617-253-1635; E-mail address: anand@mit.edu}\\
Department of Mechanical Engineering\\
Massachusetts Institute of Technology\\
Cambridge, MA 02139, USA }
\date{May 2017}
\maketitle
\section*{Abstract}

Progressive damage, which eventually leads to failure, is ubiquitous in biological and synthetic polymers. The simplest case to consider is that of elastomeric materials, which can undergo large reversible deformations with negligible rate dependence. In this paper, we develop a theory for modeling progressive damage and rupture of such materials. We extend the phase-field method, which is widely used to describe the damage and fracture of brittle materials, to elastomeric materials undergoing large deformations. A central feature of our theory is the recognition that the free energy of elastomers is not entirely entropic in nature---there is also an energetic contribution from the deformation of the bonds in the chains. It is the energetic part in the free energy which is the driving force for progressive damage and fracture.  
     
 \vspace{5mm}

\noindent Keywords: damage; rupture; polymers; flaw sensitivity; phase-field

\section{Introduction}
\label{secintro}

The development soft materials for mechanical applications has ushered a recent revolution in materials. Applications often depend on the great extensibility of polymer-based materials, as well as many other useful properties that soft materials can possess, including bio-compatibility, self-healing \citep{cordier:2008,holten-andersen:2010}, and novel actuation mechanisms and functions \citep{tokarev:2009}. In addition to the traditional engineering uses of rubbers, transformative applications are being developed daily, from surgical adhesives to replace sutures \citep{duarte:2012}, hydrogel scaffolding for tissue engineering \citep{lee:2001}, and artificial cartilage, tendons and ligaments for joint repair therapies \citep{azuma:2006,nonoyama:2016}. The mechanical demands of these applications places new importance on understanding and modeling of damage and failure of these materials. 

Modeling of failure in polymers falls in two broad categories: the first comprises macroscopic ``top-down'' approaches, based on a Griffith-type critical energy release rate criterion. The application of top-down approaches to polymers dates back to Rivlin and Thomas's work with commercial rubbers \citep{rivlin:1952}.
The second category comprises ``bottom-up'' approaches that investigate the mechanisms of damage at the molecular scale and attempt to build a consistent picture up through larger scales. 
The second approach has been largely driven by research on biological materials and design of bio-inspired composites (see, e.g., \citet{gao2006application,baer:1987,buehler2006nature,jackson1988mechanical,kamat:2000,sun2012hierarchical,sen2011structural}). An early attempt at linking these approaches occurs in the landmark paper  of \citet{Lake67Thomas}, in which they proposed a scaling law for the critical energy release rate in terms of microscopic parameters, including the binding energy between monomer units and the chain network mesh size.

One important advantage of the bottom-up approach is the understanding that it provides on the sensitivity of materials to flaws at small length scales.
The theory of Griffith governs fracture at the macro-scale, and reveals that the resistance of a body to fracture depends sensitively on the size of flaws contained within it \citep{griffith1921phenomena}. However, the picture is different at very small length scales, since the assumption that the behavior of the crack tip region can be separated from the far-field response, which underpins the Griffith theory, does not hold. Instead, 
the nonlinear mechanical response of the molecular bonds is felt over the entire body \citep{buehler2003hyperelasticity}, and a large fraction of the material is stressed to levels approaching the ideal strength, rather than just the region near the flaw tip. 
As a result, the material is much less sensitive to flaws \citep{Gao03PNAS,chen2016flaw,Mao17TalaminiEML}.
The molecular-scale physics is necessary to explain this flaw-insensitive behavior, and bottom-up modeling is necessary to control it and exploit it.

Most of the developments in the bottom-up description of failure in polymers have been conducted through the framework of molecular dynamics  \citep[see, e.g.,][]{Rottler02BarskyPRL,Rottle03RobbinsPRE}. Unfortunately, the computational demands of molecular dynamics limit the simulated length scales and time durations to scales below those needed for engineering design and optimization. It seems likely that continuum-based approaches will be needed for these purposes for the foreseeable future. Of course, to make a predictive continuum-level model, the underlying molecular physics must be retained to the furthest extent possible. The aim of this paper is to take a step in this direction.

The simplest polymers are elastomeric materials, which consist of a network of flexible polymeric chains that can undergo large reversible deformations with negligible rate-dependence.
In this paper we develop a continuum theory for modeling progressive damage and rupture of such materials.
One of the distinguishing features of elastomers is that their deformation response is dominated by changes in entropy.
Accordingly, most classical theories of rubber elasticity consider only changes in entropy due to deformation, and neglect any changes in internal energy \citep[e.g.,][]{kuhn:1942,treloar1975physics,Arruda93BoyceJMPS}.
On the other hand, as recognized by \cite{Lake67Thomas}, rupture is an \emph{energetic} process at the micro-scale, emanating from the scission of molecular bonds in the polymer chains.
In order to achieve the microscopic perspective, our model incorporates a recently proposed hyperelastic model that describes both the entropic elasticity of polymer chains, as well as a description of the mechanics of the molecular bonds in the backbone of the chain \citep{Mao17TalaminiEML}.

We make use of a phase-field approach to model the loss of stress-bearing capacity of the material due to the softening and rupture of bonds at  large stretches.
The model describes damage initiation, propagation, and full rupture in polymeric materials, and detects the transition from flaw-insensitive behavior at small scales to flaw-sensitive propagation of sharp cracks at large length scales. The phase field acts as a damage variable, with a nonlocal contribution to the free energy that regularizes the theory and sets a length scale for the rupture process.

In its structure, our framework is similar to the top-down phase field approaches to fracture based on \citet{bourdin2000numerical} that have become popular over the last 20 years, including several devoted to rupture of elastomeric materials \citep{Miehe14SchanzelJMPS,raina2016phase,wu2016stochastic}.  
The distinction of our model is in the scale: in the previous works, the phase-field serves as mathematical regularization of the critical energy release rate theory --- thus embodying a macroscopic approach --- while here we directly consider the physics of bond scission at the local scale. In particular, our work significantly departs from these previous works in the definition of the driving force for damage. 
The proposed model discriminates between entropic contributions to the free energy due to the configurational entropy of the polymer chains, and the internal energy contributions due to bond deformation. We argue that the evolution of the phase field should be driven solely by the internal energy, since the microscopic bond scission mechanism it represents is fundamentally an energetic process. 

The plan of this paper is as follows. In Section \ref{summary}, we give a brief summary of the structure of the theory, including the balance laws and the constitutive framework. In Section \ref{specialForms}, we specify the constitutive relations of the theory. We proceed step by step in the development of the constitutive relations, starting with the deformation response of a single chain with no damage, then proceed to a phase field model for softening and scission of a single chain.  We discuss how these specific constitutive relations represent a microscopic view of damage in elastomers. Next, we generalize the single chain model to describe the response of bulk material comprised of a network of chains, which bridges the microscopic perspective to the macroscopic one. In section \ref{application}, the capability of the model to describe flaw-size sensitivity of materials is illustrated. Finally, we summarize the main conclusions and make some final remarks in Section \ref{conclusions}. A full derivation of the balance laws using the principle of virtual power and the development of the thermodynamically consistent constitutive framework are included in Appendix \ref{appendix1}, and some details of the numerical implementation of the theory are given in Appendix \ref{appendix:numerical}.

\section{Summary of the constitutive theory, governing partial differential equations and boundary conditions}
\label{summary}

We have formulated a phase-field theory for fracture of a {finitely-deforming} elastic  solid using the  pioneering {virtual-power approach} of \cite{gurtin1996,gurtin2002}.    This approach leads to ``macroforce'' and ``microforce'' balances for the forces   associated with the rate-like kinematical descriptors in the theory.  
These macro- and microforce balances, together with a standard free-energy imbalance law under isothermal conditions, when supplemented with a set of thermodynamically-consistent constitutive equations, provide the governing equations for our theory.
Our theory, which is developed in detail in the Appendix, is summarized below. It relates the following basic fields:\footnote{
Notation: We use standard notation of modern continuum mechanics \citep{gurtin-fried-anand}. Specifically: $\nabla$ and Div denote the gradient and divergence with respect to the material point $\bfX$ in the reference configuration, and $\Delta =\Div \nabla$ denotes the referential Laplace operator; grad, div, and div\,grad denote these operators with respect to the point $\bfx=\bfchi(\bfX,t)$ in the deformed body; a superposed dot denotes the material time-derivative. Throughout, we write $\bfF^{e-1} = (\bfF^e){}^{-1}$, $\bfF^{e-\trans}=(\bfF^e)^{-\trans}$, etc. We write $\Tr\A$, $\sym\A$, $\skw\A$, $\A_0$, and $\symz\A$ respectively, for the trace, symmetric, skew, deviatoric, and  symmetric-deviatoric parts of a tensor $\A$. Also, the inner product of tensors $\A$ and $\bfB$ is denoted by $\A \tendot \bfB$, and the magnitude of $\A$ by $|\A|=\sqrt{\A \tendot \A}$.
\label{notation}}

\begin{center}

\begin{tabular}{ll}
 $\bfx=\bfchi(\bfX,t) $, &  motion;\\[2pt]
 $\bfF=\nabla\bfchi, \quad J=\det\bfF>0$, &    deformation gradient;\\[2pt]
$ \Fbar = J^{ \,-1/3}\,\F$, &  disortional part of $\F$;\\[2pt]
 $\C=\FT\F$,   &   right  Cauchy-Green tensor;\\[2pt]
 $\Cbar= \Fbar^{\trans}\Fbar=J^{-2/3}\C$, &distortional part of $\C$;\\[2pt]
%
%
 %
%
%
%
%
%
%
$\bfT_\mat$, $\bfT_\mat \F^{\trans} = \F \Tmat^{\trans} $ & Piola stress; \\[2pt]
$\T_{\mat\mat} =\F^{-1} \bfT_\mat$,& second Piola stress;\\[2pt]
$\psi_\mat $, & free energy density per unit reference volume; \\[2pt]
$\varepsilon_\mat$, & internal energy density per unit reference volume; \\[2pt]
 $\lambda_b>0$ &   effective bond stretch (an internal variable);\\[2pt]
$d(\X,t) \in [0,1]$, &   order parameter, or phase-field;\\[2pt]
 $\varpi$ &   scalar microstress;\\[2pt]
 $\bfxi$ &    vector microstress.
\end{tabular}
\end{center}

\subsection{Constitutive equations}

 \begin{enumerate}
 \setlength{\itemsep}{6pt}

 \item {\bf Free energy}
 
This is given by
 \begin{equation} 
 \psi_\mat = \, \hat{\psi}_\mat( \bfLambda),
  \label{summ1}
 \end{equation}
  with $\bfLambda$ the list
\begin{equation}
\bfLambda =\{\C,\lambda_b,d,\nabla d \}.
\label{summ1a}
\end{equation}

 \item {\bf Second Piola stress.    Piola stress. Cauchy stress. Kirchhoff stress}

   The  second Piola  stress is given by
     \begin{equation}
   \T_{\mat\mat}  =   2\, \pards{\hat{\psi}_\mat(\bfLambda)}{\C},
\label{summ2a}
\end{equation}
and the Piola stress by   
   \begin{equation}
    \T_\mat =  \F\T_{\mat\mat}.
    \label{summ4b}
\end{equation}

\item{\bf Implicit equation for the effective bond stretch}

The thermodynamic requirement 
\begin{equation}
\pards{\hat{\psi}_\mat(\bfLambda)}{\lambda_b}=0,
     \label{summ4d}
\end{equation}
serves as an implicit equation to determine  the effective bond stretch $\lambda_b$, in terms of the other constitutive variables.

\item{\bf Microstresses  $\varpi$   and $\bfxi$}

The scalar microstress $\varpi$  is  given by
\begin{equation}
    \begin{aligned}
        \varpi&= \underbrace{\pards{\hat{\psi}_\mat(\bfLambda)}{\dee}}_{\varpi_\en}  +\underbrace{\alpha + \zeta \deedot}_{ \varpi_\diss}
    \end{aligned}
\label{summgenac1a}
\end{equation}
with  $\alpha=\hat\alpha(\bfLambda)$  and $\zeta=\hat\zeta(\bfLambda)$ positive-valued scalar functions.
Here $\varpi_\en$ and $\varpi_\diss$ denote the energetic and dissipative parts of $\varpi$.

The vector microstresses   $\bfxi$ is given by,
\begin{equation}
    \begin{aligned}
        \bfxi &=  \pards{\hat{\psi}_\mat(\bfLambda)}{\nabla\dee},
    \end{aligned}
\label{summgenac1}
\end{equation}
and is taken to be energetic, with no dissipative contribution.
\end{enumerate}

\subsection{Governing partial differential equations}

The governing partial differential equations consist of
\begin{enumerate}
\item
\textbf{Equation of motion}:  

\begin{equation}
\fbox{ \parbox{4cm} {\[
 \Div \T_\mat  +    \bfb_{0 \mat} = \rho_\mat \ddot\y,
  \]}} 
 \label{macfb2}
\end{equation}
where $\bfb_{0\mat}$ is a non-inertial body force,  $\rho_\mat$ is the referential mass density, $\ddot\y$ the acceleration, and the Piola stress  $\T_\mat$  is given by \eqref{summ4b}.
\item \textbf{Microforce balance}: 

The microforces $\varpi$ and  $\bfxi$  obey the  balance  \eqref{vpower28}, viz.
\begin{equation}
\Div \bfxi -\varpi=0.
\label{mfbsunn}
\end{equation}
This microforce balance, 
together with the thermodynamically consistent constitutive equations \eqref{summgenac1a} and \eqref{summgenac1} for $\varpi$ and $\bfxi$ gives the following
evolution equation for the  damage variable $\dee$,\footnote{We use the phrases ``order parameter'', ``phase-field'', and ``damage variable'' interchangeably to describe $\dee$.}
\begin{equation}
\fbox{ \parbox{8cm} {\[
    \begin{aligned}
        \hat{\zeta} (\bfLambda) \dot{d}  & = 
        -\pards{\hat{\psi}_\mat(\bfLambda)}{d} + \Div \left( \pards{\hat{\psi}_\mat(\bfLambda)}{\nabla d} \right) - \hat\alpha(\bfLambda).
    \end{aligned}
 \]}} 
\label{summgenac}
\end{equation}
Since $\zeta$ is positive-valued,  the right hand side of \eqref{summgenac} must be  positive for $\deedot$ to be positive and the damage to increase monotonically.

\end{enumerate}

\subsection{Boundary and initial conditions}
\label{boundaryconds}

We also need boundary and initial conditions to complete the theory.  

\begin{enumerate}
\item \textbf{Boundary conditions for the pde governing the evolution of $\y$}: 

Let $\calS_{\bfchi}$ and $\calS_{\bft_\mat}$ be  {complementary subsurfaces} of the boundary $\partial\B$ of the body $\B$.
%
Then for  a time interval $t\in[0,T]$ we consider a pair of boundary conditions in which the motion is specified on $\calS_{\bfchi}$ and the surface traction on $\calS_{\bft_\mat}$:
\begin{equation}
\left.
\begin{aligned}
\bfchi=\breve\bfchi \quad &\text{on $\calS_{\bfchi} \times [0,T]$},
\\[4pt]
\T_\mat\bfn_\mat=\breve{\bft}_\mat\quad &\text{on $\calS_{\bft_\mat}  \times [0,T]$}.
\end{aligned}
\right\}
\label{staticbc1}
\end{equation}
In the boundary conditions above  $\breve\bfchi$ and $\breve\bft_\mat$ are \emph{prescribed} functions of $\bfX$ and $t$.

\item \textbf{Boundary conditions for the pde governing the evolution of $d$}:

The presence of microscopic stresses $\bfxi$ results in an expenditure of
power 
\[
    \int_{\partial{}\B}(\bfxi \!\cdot\bfn_\mat) \dot{d} \; da_\mat
\]
by the material in contact with the body (cf., \eqref{vpower29}), and this necessitates a
consideration of boundary conditions on $\partial{}\B$ involving
the microscopic tractions $\bfxi\!\cdot\bfn_\mat$ and the 
rate of change of the damage variable $\dot{d}$.  
\begin{itemize}
\item 
\emph{We restrict attention to boundary conditions that
result in a null expenditure of microscopic power in the sense
that $(\bfxi\!\cdot\bfn_\mat) \dot{d} = 0$}. 
\end{itemize}
A  simple set of  boundary conditions which satisfies this requirement is, %
\begin{equation}
    \begin{aligned}
        \dot{d} = 0 \quad &\text{on $\calS_{d} \times [0,T]$},\\
        \bfxi\cdot\bfn_\mat = 0 \quad &\text{on $\partial\B\setminus\calS_{d}  \times [0,T] $},
    \end{aligned}
\label{pde8}
\end{equation}
with the microforce $\bfxi$ given by \eqref{summgenac1}. 

\end{enumerate}
 The initial data is taken as 
\begin{equation}
\bfchi(\X,0)=\X, \quad \quad \dot\bfchi(\X,0)=\bfv_0(\X), \quad \textrm{and} \quad
  d(\X,0)=0.  \quad \textrm{in} \quad \B.
\label{hcibvp31}
\end{equation}

The coupled set of equations \eqref{macfb2} and  \eqref{summgenac}  together with  \eqref{staticbc1}, \eqref{pde8}, and \eqref{hcibvp31}  yield an initial/boundary-value problem for the motion $\bfchi(\bfX,t)$, and the phase-field $d(\X,t)$.

 \section{Specialization of the constitutive equations}
\label{specialForms} 

As stated in the introduction, we wish to characterize the process of rupture in elastomeric material in terms of the microscopic mechanics of molecular bond scission between the backbone units of the polymer chains. However, the traditional hyperelastic models for elastomers neglect the energetics of bond deformation. In order to make this connection between the scales, we make use of our recently proposed hyperelastic constitutive model that accounts for the energetics of bond deformation, as well as the well-known entropic elasticity effects in polymers \citep{Mao17TalaminiEML}. 
In what follows we begin by considering the process of deformation, damage, and rupture of a single chain, and then extend the single chain considerations to bulk elastomeric materials.

\subsection{Deformation, damage, and fracture  of a single chain }
\label{section:single_chain}

\subsubsection{Free energy of a single chain  in the absence of damage}

\begin{figure}[htbp]
    \centering
    \includegraphics[scale=1]{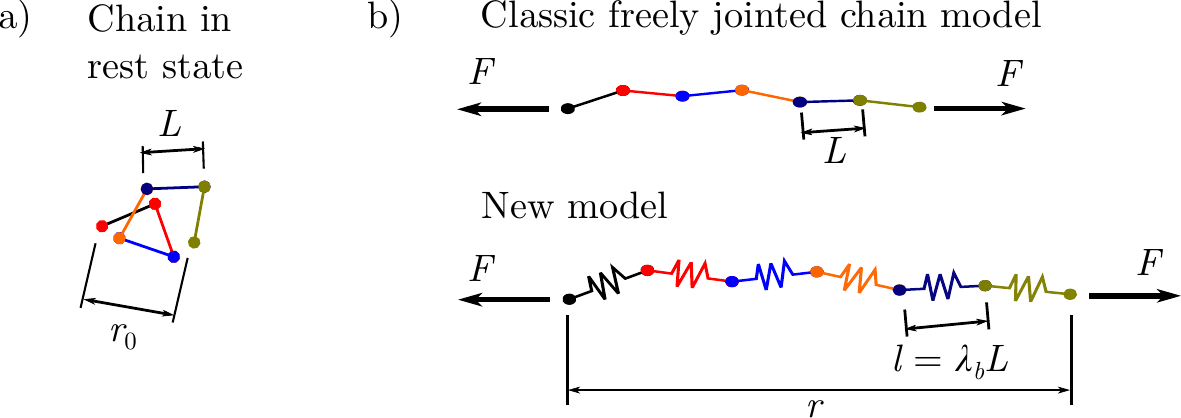}
    \caption{a) Chain in the rest state. The Kuhn segment length in the rest state is $L$. b) Upper panel: classic freely jointed chain model under stretch. The Kuhn segments are assumed rigid. Bottom panel: model of \citet{Mao17TalaminiEML}. The Kuhn segments are assumed deformable due to deformation of the constituent bonds.}
    \label{fig:bond_stretch_schematic}
\end{figure}

To illustrate the model of \citet{Mao17TalaminiEML}, let us consider first the behavior of a single chain. We make the kinematic assumption that the overall deformation of the polymer chain under load is due to two sources, (i) the alignment of the Kuhn segments in the chain under load, and (ii) stretching of the segments due to deformation of the constituent molecular bonds (see Figure \ref{fig:bond_stretch_schematic}). 

Consider a single chain with $n$ segments, each of initial length $L$, and as is standard, let $r_0 = \sqrt{n}L$  denote the unstretched chain length determined from random walk statistics. 
With $r$ denoting the end-to-end distance of chain in a deformed configuration, let $\lambda = r/r_0$ denote the overall chain stretch.
The current segment length $l$  is related  to the rest length $L$ through $l = L \lambdab$, where $\lambdab$ is a dimensionless stretch which we refer to as the \emph{bond deformation stretch}. 
Using the Langevin statistics for the chain segments as developed in \citet{kuhn:1942}, the entropy of the chain is \citep{Mao17TalaminiEML},
\begin{align}
    \eta &= \hat{\eta} (\lambda, \lambdab)
    = -n k_B \left[ \frac{\lambda \lambdab^{-1}}{\sqrt{n} } \beta + \ln\left(\frac{\beta}{\sinh \beta}\right) \right], 
    \label{entropy}
\end{align}
with
\begin{align}
    \beta  = \mathcal{L}^{-1}\left( \frac{\lambda \lambdab^{-1} }{\sqrt{n} } \right),
    \label{entropy2}
\end{align}
where $\mathcal{L}^{-1}$ denotes the inverse of the Langevin function $\mathcal{L}(x) = \coth x - x^{-1}$, and $k_B$ is Boltzmann's constant.
Note from \eqref{entropy} and \eqref{entropy2},  that it is  
\begin{equation}
\text{ the modified stretch measure $(\lambda \lambda_\text{\tiny b}^{-1})$  which gives rise to changes in the entropy of the chain}.
    \label{entropy3}
\end{equation}
We refer to $(\lambda \lambda_\text{\tiny b}^{-1})$ as the \emph{stretch due to segment realignment}.

Next, we consider how the deformation of the bonds causes the internal energy of the chain $\varepsilon$ to change.
In the following, we use the simple functional form for the internal energy of the chain,
\begin{equation}
    \varepsilon = \hat{\varepsilon} (\lambda_b) = \frac{1}{2} n E_b (\lambda_b - 1)^2,
    \label{internal_energy}
\end{equation}
where $E_b$ is a parameter with units of energy that characterizes the bond stiffness. 

The Helmholtz free energy density is defined as $ \psi = \varepsilon - \thet \eta$,
which, upon substituting the specialized constitutive equations \eqref{entropy} and \eqref{internal_energy} yields,
\begin{align}
    \psi =  
    \hat{\psi} (\lambda, \lambdab) &= 
    \frac{1}{2} n E_b \left( \lambdab - 1 \right)^2
    +n k_B \thet \left[ \left(\frac{\lambda \lambdab^\inv}{\sqrt{n}}\right) \beta
    + \ln\left(\frac{\beta}{\sinh \beta}\right)\right].
    \label{eq:undamaged_free_energy}
\end{align}

Next, to find $\lambdab$, we use the fact that at fixed $\lambda$, a particular value of $\lambda_b$ will minimize the free energy (cf., \eqref{summ4d}), and this will be the most probable state in which to find the system. Thus,
\begin{equation}
    \lambda_b = \arg \min_{\lambda_b^{*} > \lambda/\sqrt{n}} \hat{\psi}_\mat(\lambda, \lambda_b^{*}),
    \label{bond_stretch_condition}
\end{equation}
which provides an implicit, nonlinear equation to determine $\lambdab$.

Note that in the classical freely jointed chain model there is no bond stretch, i.e., $\lambda_b=1$, and there is no internal energy contribution to the free energy $\psi$. In this case, as $\lambda \to \sqrt{n}$ the entropy vanishes and the free energy \emph{diverges}.
In contrast, the bond deformation mechanism \eqref{internal_energy} in the free energy expression \eqref{eq:undamaged_free_energy} ensures that the quantity $(\lambda\lambda_b^\inv)$ is always less than than $\sqrt{n}$ and that the free energy \emph{remains finite}.

\subsubsection{Accounting for damage and scission of a single chain}

For simplicity,  we assume that thermal effects are negligible and that all bonds in the chain   stretch uniformly,  and eventually damage and   fail under increasing stretch. 
The softening is due to the weakening of the molecular attraction between monomer units as they are separated. 
Thus, introducing a damage variable  $d \in [0,1]$, this assumption leads us to adopt the following  functional form for the free energy of a single chain,
\begin{equation}
    \hat{\psi} (\lambda, \lambda_b, d) = g(d) \hat{\varepsilon}^0 (\lambda_b) - \vartheta \hat{\eta}(\lambda, \lambda_b),
    \label{free_energy_chain}
\end{equation}
where $\hat{\varepsilon}^0( \lambda_b)$ is the internal energy of an \emph{undamaged} chain, as given by \eqref{internal_energy}, and the function $g(d)$ is a monotonically decreasing \emph{degradation function} with value,
\[
    g(0)=1,
\]
which produces the usual elastic behavior of the bonds in the intact state $d=0$, and has a value 
\[
    g(1) = 0,
\]
to represent the fully damaged state. In other words, the degradation function describes the decreasing stiffness of the bonds under large stretches.
Additionally, the degradation function is subject to the constraint
\[
    g'(1) = 0,
\]
so that the thermodynamic driving force for damage $\varpien = \partial \hat{\psi} / \partial d$ vanishes as the chain becomes fully damaged.\footnote{cf.  eq. \eqref{summgenac1a}  for definition $\varpien$.}
A widely-used degradation function is 
\begin{equation}
    g(d)=(1-d)^2;
    \label{specialized_deg}
\end{equation}
we adopt it here.\footnote{In numerical calculations the degradation function is modified as
$$
g(d)=(1-d)^2 +k,
$$
where $k$ is a small, positive-valued constant, which is  introduced to prevent ill-conditioning of the model when $d=1$. \label{foot:regularization}}

Consistent with our treatment of the segments as having equal stretch, the total dissipation from scission is approximately the binding energy of the monomer units times the number of monomer units in the chain. To reflect this in the model, we set the rate-independent  part  $\alpha$ of the dissipative microstress  $\varpidiss$ to\footnote{cf. eq. \eqref{summgenac1a}  for definition $\varpidiss$.} 
\begin{equation}
    \alpha = n \varepsilon_b^f,
    \label{chain_diss_microforce}
\end{equation}
where $\varepsilon_b^f$ is the binding energy between the monomer units.

In the present considerations for a single chain, where the $\nabla d$ term does not come into play, the microforce balance \eqref{summgenac} becomes 
\begin{equation}
    \zeta \dot{d} = -\pards{\hat{\psi}(\lambda, \lambda_b, d)}{d} - \alpha.
    \label{chain_microforce_balance}
\end{equation}
Let us consider the rate independent limit ($\zeta = 0$). Then, inserting \eqref{free_energy_chain} and \eqref{specialized_deg} into this balance, and using the fact that  $\dee$ lies in the range $d \in [0,1]$, we have that 
\begin{equation}
    d =
    \begin{cases}
        0, & \text{if } \hat{\varepsilon}(\lambda_b) \leq n \varepsilon_b^f / 2,
        \\
        1 - \dfrac{n \varepsilon_b^f / 2}{\hat{\varepsilon}^0(\lambda_b)} & 
        \text{if }
        \hat{\varepsilon}(\lambda_b) > n \varepsilon_b^f / 2.
    \end{cases}
    \label{local_phasefield_evolution}
\end{equation}
The bond deformation stretch $\lambda_b$ is determined implicitly by \eqref{summ4d}, which yields the nonlinear equation
\begin{equation}
    (1-d)^2 E_b (\lambda_b -1) - k_B \vartheta \frac{\lambda}{\sqrt{n} \lambdab^2} \mathcal{L}^{-1}\left( \frac{\lambda\lambdab^\inv}{\sqrt{n}}  \right) = 0.
    \label{chain_bond_stretch_condition}
\end{equation}
Given an imposed chain stretch $\lambda$, equations \eqref{local_phasefield_evolution} and \eqref{chain_bond_stretch_condition} can be solved simultaneously for $d$ and $\lambda_b$.

\begin{figure}[htbp]
    \centering
    \begin{subfigure}[b]{0.48\textwidth}
        \includegraphics[width=1.0\textwidth]{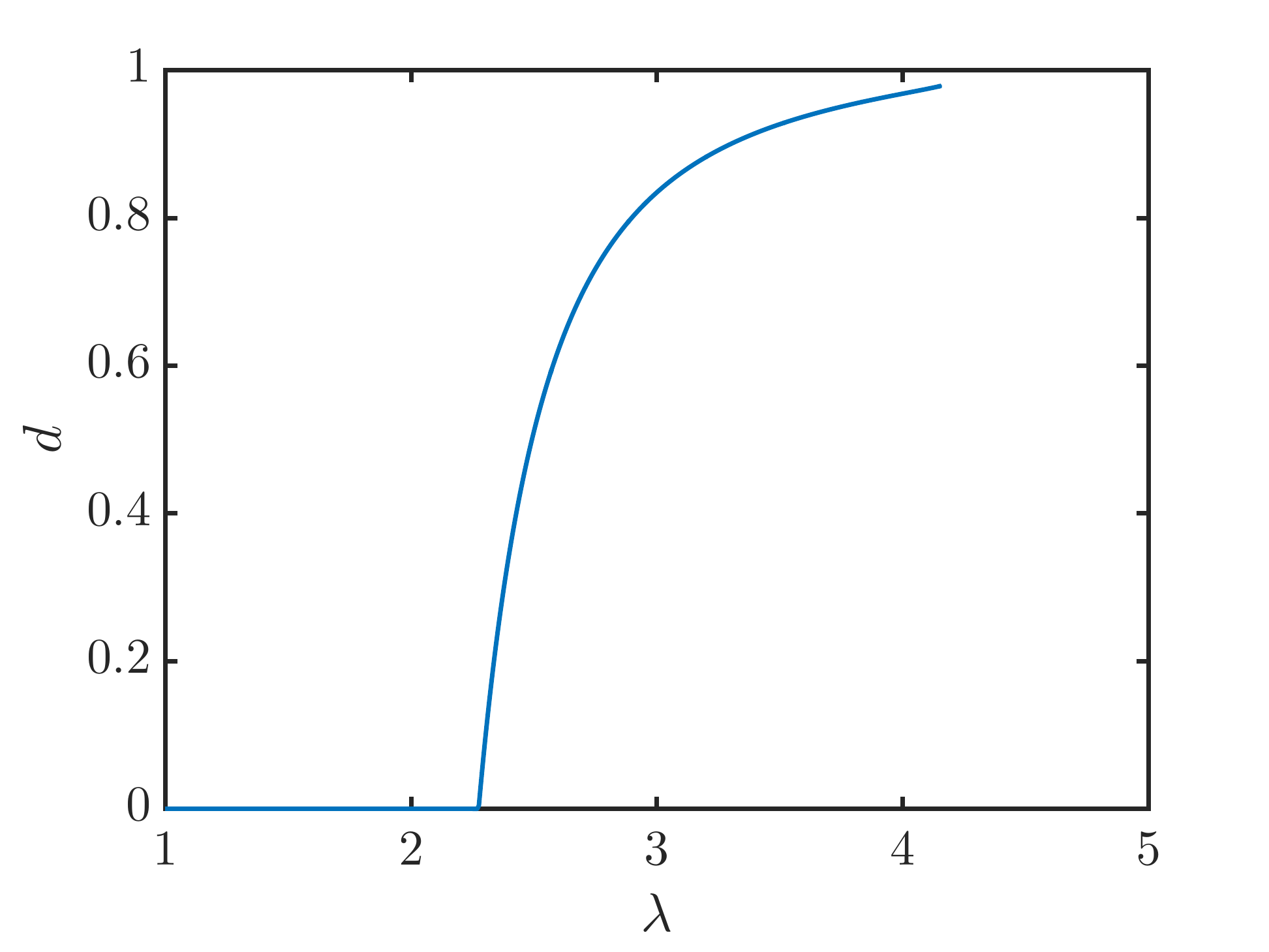}
        \caption{ }
    \end{subfigure}
    \begin{subfigure}[b]{0.48\textwidth}
        \includegraphics[width=1.0\textwidth]{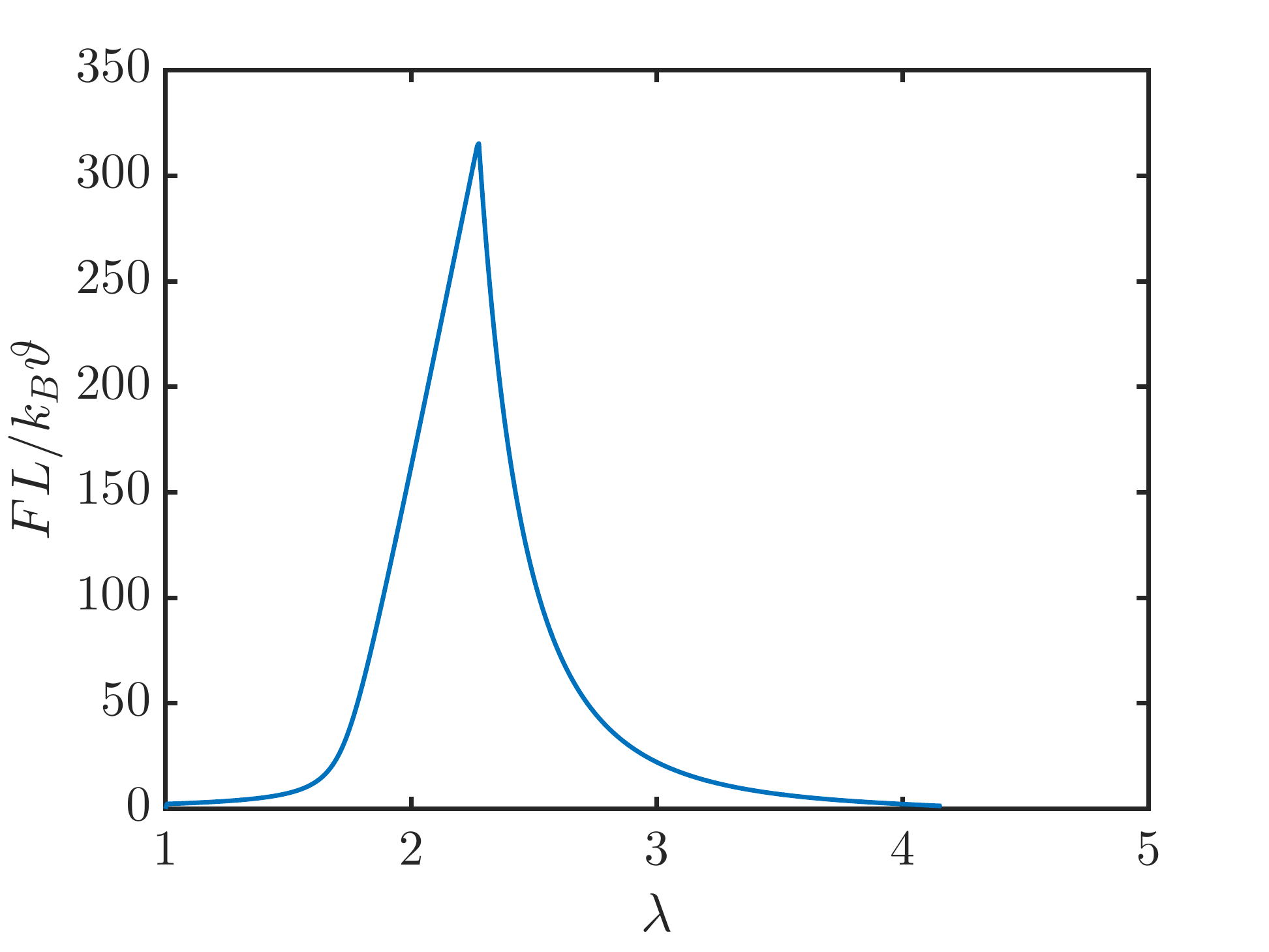}
        \caption{ }
    \end{subfigure}
    \caption{Single chain response. (a) Evolution of the damage variable $d$ with chain stretch $\lambda$. (b) Normalized chain force $FL / k_B \vartheta$ versus chain stretch.}
    \label{chain_force}
\end{figure}

We may visualize the behavior of the model with the following simple example, in which we take 
\begin{align*}
    n = 3, && E_b / k_B \vartheta = 1000, && \varepsilon_b^f / k_B \vartheta = 100,&&
\end{align*}
and impose a monotonically increasing stretch and sketch the resulting response.\footnote{
    We have intentionally chosen a small value $n=3$ for the number of links in the chain to illustrate the features of our theory so that failure of the chains in our  simulations occurs at reasonable levels of stretch $\lambda$.
}
The evolution of the damage variable $d$ is plotted against the imposed stretch $\lambda$ in Figure \ref{chain_force}a. The damage variable $d$ remains zero until the internal energy reaches the critical value $\varepsilon_b^f / 2$ (which occurs at $\lambda \approx 2.3$). After reaching this point, the damage variable increases and asymptotically approaches unity according to \eqref{local_phasefield_evolution}. In figure \ref{chain_force}b, the chain force $F$ (presented in the dimensionless form $F L / k_B \vartheta = (\sqrt{n} k_B \vartheta)^{-1} \partial \hat{\psi} / \partial \lambda$) is plotted against the imposed stretch $\lambda$. The force follows the undamaged stiffening response until the critical value of the internal energy is reached. At that point, the damage variable begins to increase according to \eqref{local_phasefield_evolution}, and the bond stiffness begins to degrade, leading to a decrease in force with increasing stretch.

\begin{figure}[htbp]
    \centering
    \begin{subfigure}[b]{0.48\textwidth}
        \includegraphics[width=1.0\textwidth]{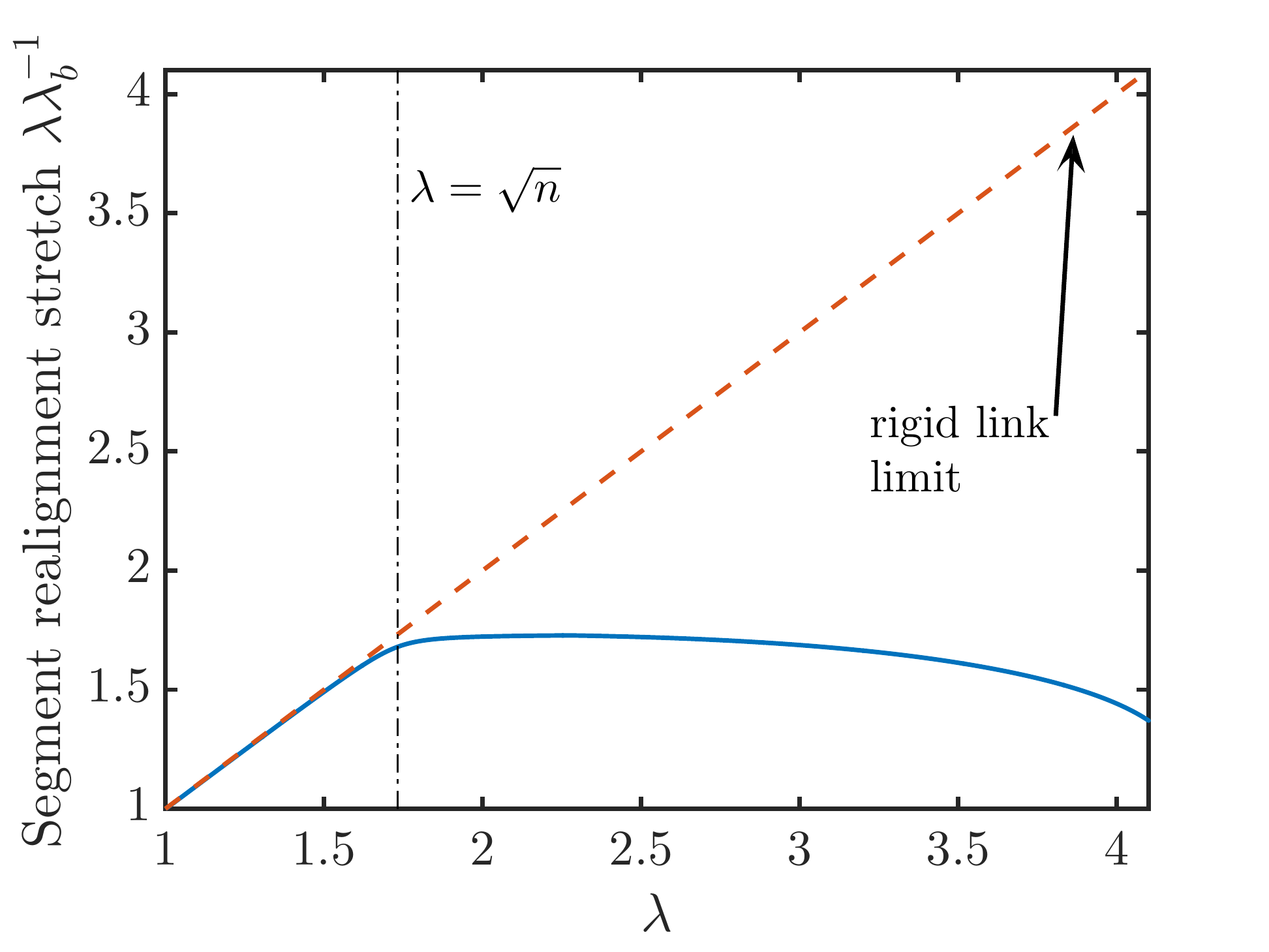}
        \caption{ }
    \end{subfigure}
    \begin{subfigure}[b]{0.48\textwidth}
        \includegraphics[width=1.0\textwidth]{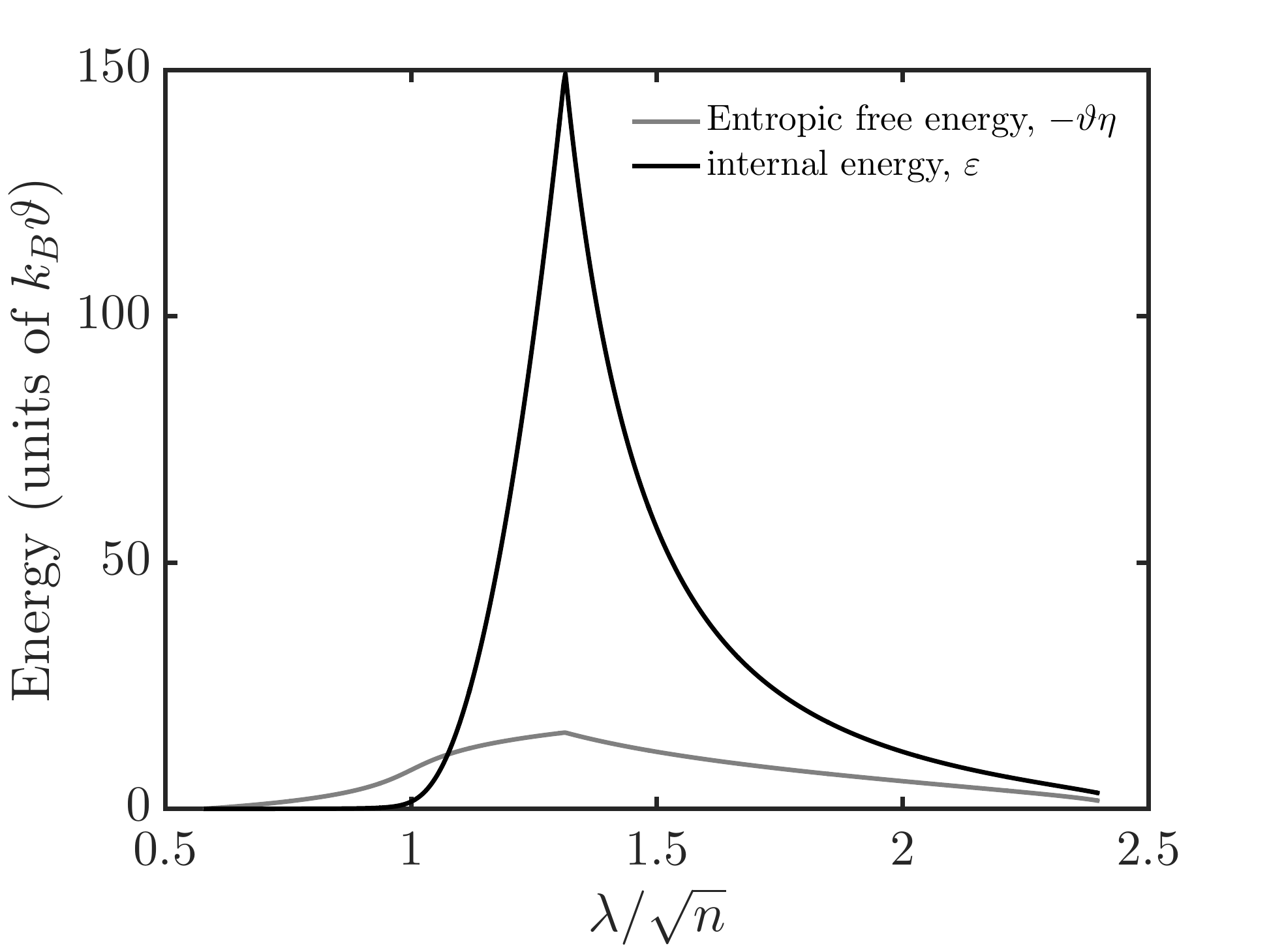}
        \caption{ }
    \end{subfigure}
    \caption{Single chain model response. (a) Evolution of entropic stretch $(\lambda\lambda_b^\inv$ with chain stretch $\lambda$. (b) Variation of the  contributions $(-\thet\eta)$ and $\varepsilon$  to the free energy with chain stretch.}
    \label{free_energy_components}
\end{figure}

Next, we illustrate the deformation mechanisms in the model. Recall from \eqref{entropy3} that the stretch measure $(\lambda \lambda_b^{-1})$  gives rise to the change in the entropy of the chain.  Figure \ref{free_energy_components}a  shows a  plot of  $\lambda \lambda_b^{-1}$ against the imposed stretch $\lambda$. When the imposed stretch is below $\sqrt{n}$ (the limiting stretch in the classical freely jointed chain model), the overall chain stretch and the stretch due to segment realignment are virtually identical. In other words, the behavior of the chain in this range is virtually identical to the classic freely jointed chain model, which has perfectly rigid links.
Since the bonds are quite stiff in this example ($E_b / k_B \vartheta = 1000)$, it is energetically favorable for the system to accommodate stretch primarily through the entropic mechanism when $ \lambda < \sqrt{n}$. This is further demonstrated in Figure \ref{free_energy_components}b, where both components of the free energy, entropic $(-\vartheta \eta)$ and the internal energy $(\varepsilon)$, are plotted. The internal energy due to bond stretching is seen to be negligible until $\lambda \approx \sqrt{n}$. On the other hand, as the stretch passes $\sqrt{n}$, the chain approaches its contour length, the configurational entropy is nearly exhausted, and it becomes energetically favorable to accommodate additional stretch through bond deformation. Thus, the realignment stretch $(\lambda \lambda_b^{-1})$ remains approximately constant as the overall stretch $\lambda$ increases beyond $\sqrt{n}$ in Figure \ref{free_energy_components}a. 

The next major event occurs when the bonds reach the critical energy for softening to begin ($\lambda \approx 2.3 \approx 1.3 \sqrt{n}$). To visualize the response, one may imagine the system as a mechanical model of two springs in series, one representing the entropic elasticity, and the other representing the energetic elasticity.
The parameter $(1-d)^2 E_b$ can be considered the effective stiffness of the energetic elasticity spring. As the damage variable increases, $(1-d)^2 E_b$ decreases, and the stretch of the energetic spring increases at the expense of the stretch in the entropic spring. In other words, the bond deformation stretch $\lambda_b$ increases and the segment realignment stretch $(\lambda \lambdab^{-1})$ \emph{decreases} once damage begins, as seen in Figure \ref{free_energy_components}a. It follows that during damage, the entropy begins to \emph{increase}. This is expected, as the length of the Kuhn segments is rapidly increasing, and with longer Kuhn segments there are a greater number of configurations possible to achieve the overall stretch. 

\begin{Remark}
The microforce balance \eqref{chain_microforce_balance} can be rewritten to enforce the constraint $d \in [0,1]$, which leads to \eqref{local_phasefield_evolution}  in a simple way. 
To find it, first substitute \eqref{free_energy_chain}, \eqref{specialized_deg}, and \eqref{chain_diss_microforce} into the microforce balance \eqref{chain_microforce_balance}, and then add and subtract the term $n \varepsilon_b^f d$ to get
\begin{equation*}
    \zeta \dot{d} = 2(1-d) \left( \hat{\varepsilon}^0 (\lambdab) - n \varepsilon_b^f/2 \right) - n \varepsilon_b^f d.
\end{equation*}
The constraint is automatically satisfied if the equation  above is modified to read as,
\begin{equation}
    \zeta \dot{d} = 2(1-d) \left\langle \hat{\varepsilon}^0 (\lambdab) - n \varepsilon_b^f/2 \right\rangle + n \varepsilon_b^f d,
    \label{chain_microforce2}
\end{equation}
where $\langle \bullet \rangle$ are Macauley brackets, i.e.,
\[
    \langle x \rangle =
    \begin{cases}
        0, & x < 0,
        \\
        x, & x \geq 0.
    \end{cases}
\]
In this form, the threshold for the damage driving force is made explicit.
\end{Remark}

\subsection{Deformation, damage, and fracture of a network of chains}
\label{network_theory}

We employ the widely-used eight-chain network representation of \citet{Arruda93BoyceJMPS} to extend the single chain model to a continuum model.\footnote{
Note that in addition to the usual assumption of weak chain interactions, we must additionally assume that the damage in each chain is independent in order apply the Arruda-Boyce  network model.}
The entropy and energy of the network can be obtained by summing the contributions from individual chains as given by the single chain model.
To this end, we follow \citet{anand1996constitutive} and define the effective chain stretch
\begin{equation}
    \lambdabar \Def \sqrt{\Tr \Cbar/3},
\end{equation}
where $\Cbar$ is the distortional right Cauchy-Green tensor.  With $N$ representing the number of chains per unit volume  of the reference configuration, 
the entropy density of the network is then given by
\begin{equation}
    \eta_\mat = \hat{\eta}_\mat (\lambdabar, \lambda_b)
    = - N k_B n \left[ \left(\frac{\lambdabar \lambda_b^\inv}{\sqrt{n}}\right) \beta + \ln\left(\frac{\beta}{\sinh \beta}\right)  \right],\qquad \beta = \mathcal{L}^{-1} \left( \frac{\lambdabar \lambda_b^\inv}{\sqrt{n}} \right).
    \label{network_entropy}
\end{equation}
For the internal energy density of a network we allow for a dependence on $\lambda_b$ and $\dee$ as in our consideration for a  single chain, but here we also allow for internal energy contribution due to volumetric changes, $J$,  and also a dependence on the gradient of the damage, $\nabla \dee$:
\begin{equation}
    \varepsilon_\mat = \hat{\varepsilon}_\mat(\lambda_b, J, d, \nabla d) =
    g(d) \hat{\varepsilon}_\mat^0 (\lambdab, J)
    + \hat{\varepsilon}_{\mat, \text{nonloc}} (\nabla d).
    \label{network_internal_energy}
\end{equation}
For the undamaged part of the internal energy $ \hat{\varepsilon}_\mat^0 (\lambda_b, J)$  we  choose the constitutive relation
\begin{equation}
    \hat{\varepsilon}_\mat^0 (\lambda_b, J) =    \dfrac{1}{2} N\, n\,E_b   (\lambda_b - 1)^2 
    + \half K(J-1)^2,
\end{equation}
where the first term represents the internal energy of bond stretching, and the second term models the slight compressibility of the material, with $K$ the bulk modulus.\footnote{
We have encountered some  difficulties with this form of the volumetric internal energy in our  finite element simulations. At late stages of the damage, the shear stiffness degrades faster than the volumetric stiffness, and the near-incompressibility constraint becomes severe. The particular form of the volumetric part of the internal energy is not crucial in elastomeric materials, where volume changes are typically quite small relative to distortional deformations. Accordingly, we have used the form $\frac{K}{8} (J-1/J)^2$ \citep[see][]{schroderInvariant} for the volumetric internal energy in computations, which leads to a softer response at large $J$. }

The term $\hat{\varepsilon}_{\mat, \text{nonloc}} (\nabla d)$ in the internal energy density is the nonlocal contribution
\begin{equation}
    \hat{\varepsilon}_{\mat, \text{nonloc}}(\nabla d) = \dfrac{1}{2}\kappa \left| \nabla d \right|^2,
    \label{nonlocalterm}
\end{equation}
where $\kappa$ is a parameter with dimensions of energy per unit length. The quantity
\begin{equation}
      \varepsilon_\mat^f \Def N n \varepsilon_b^f.
    \label{epsfdef}
\end{equation}
represents the energy of chain scission per unit volume when all bonds are broken. 
We may use \eqref{epsfdef} to  express $\kappa$ as
\begin{equation}
    \kappa = \varepsilon_\mat^f \, \ell^2,
    \label{elldef}
\end{equation}
where $\ell$ represents an intrinsic length scale for the damage process.
%
The nonlocal  term \eqref{nonlocalterm} penalizes steep gradients in the damage variable $\dee$. Physically, the zone of damage due to chain scission cannot be smaller than the mean distance between cross-links, which is the only natural length scale of the network structure. Even though the damage criterion is based directly on the local chain scission energetics, the nonlocal term also ensures that the model predicts a well-defined critical energy release rate in the macroscopic limit.\footnote{
This is illustrated through an example in the Section \ref{energy_release_rate}.}

\begin{Remark}
A macroscopic critical energy release rate $G_c$ can be estimated for the case of strongly bonded elastomers. In this scenario, the internal energy will significantly outweigh the entropic free energy at the point of scission, and thus the entropic free energy contribution to the energy release rate is negligible. In the phase field model, the dissipation scales as $\varepsilon_\mat^f \ell^3$, while for a theoretical sharp crack it would scale as $G_c \ell^2$, so that we must have
\begin{equation*}
    G_c \ell^2 \sim \varepsilon_\mat^f \ell^3.
\end{equation*}
Rearranging this result yields
\begin{equation}
    \ell \sim \frac{G_c}{n (N k_B \vartheta)} \left(\frac{\varepsilon_b^f}{k_B \vartheta} \right)^{-1},
\end{equation}
which provides a means for estimating the parameter $\ell$ in terms of the macroscopically measurable parameters $G_c$, $n$, $N k_B \vartheta$  (the ground state shear modulus), and the binding energy $\varepsilon_b^f$, which is tabulated for commonly occurring repeat unit bonds.
\end{Remark}

With the constitutive relations \eqref{network_entropy}-\eqref{elldef} for $\varepsilon_\mat$ and $\eta_\mat$ in hand, the free energy follows from the identity
\begin{equation}
    \psi_\mat = \varepsilon_\mat - \vartheta \eta_\mat.
\end{equation}

To complete the specification of the constitutive relations, we specify the dissipative microforce $\varpidiss$ that expends power through $\dot{d}$.\footnote{
Cf.  eq. \eqref{summgenac1}.}
The dissipative microforce is partitioned into a rate independent part and a rate dependent part through
\begin{equation}
    \varpidiss = \underbrace{\alpha}_\text{rate independent} + \underbrace{\zeta \dot{d}}_\text{rate dependent}.
\label{varpispec1}
\end{equation}
The rate-independent part of the dissipative microforce $\alpha$ is the sum of the contributions from each chain given by \eqref{chain_diss_microforce}, thus
\begin{equation}
    \alpha = \varepsilon_\mat^f.
    \label{varpispec2}
\end{equation}
%
The rate-dependent contribution to the dissipative microforce $\zeta \dot d$, is simply described by a  constant kinetic modulus $\zeta > 0$, with the rate-independent limit of damage evolution given by $\zeta \to 0$.

Using the specializations above, the microforce balance \eqref{summgenac}, which gives the evolution of $\dee$, becomes
\begin{equation}
    \begin{split}
        \zeta \dot{d} 
&= 2(1-\dee)\hat\varepsilon_\mat^0(\lambdab,J)  +  \varepsilon_\mat^f\ell^2 \Delta\dee -\varepsilon_\mat^f.
    \end{split}
\label{summzz1}
\end{equation}
Keeping in mind the threshold for damage intiation  \eqref{local_phasefield_evolution}, and to make connection with the previous work of Miehe and co-workers \citep[cf., e.g.,][]{miehe2010,Miehe14SchanzelJMPS}, we apply the same technique used in \eqref{chain_microforce2} to rewrite the evolution equation \eqref{summzz1} as
\begin{equation}
    \zeta \dot{d} 
    = 2(1-\dee) \left\langle \hat\varepsilon_\mat^0(\lambdab,J) -  \varepsilon_\mat^f/2 \right\rangle 
    - \varepsilon_\mat^f\left(\dee - \ell^2 \Delta\dee\right).
    \label{summzz2}
\end{equation}


At this stage, the irreversible nature of scission is not yet reflected in the model. To this end, we replace the term $\langle \hat{\varepsilon}_\mat^0 (\lambdab, J) - \varepsilon_\mat^f/2\rangle$ in the microforce balance with the monotonically increasing history field function  \citep[cf.,][]{miehe2010}:
\begin{equation}
    \mathcal{H}(t) \Def \max_{s \in [0,t]} 
    \left\langle \hat{\varepsilon}_\mat^0 (\lambdab(s), J(s)) - \varepsilon_\mat^f/2
    \right\rangle.
\end{equation}
The microforce balance \eqref{summgenac} then becomes
\begin{equation}
    \zeta \dot{d} = 2 (1-d) \mathcal{H} - \varepsilon_\mat^f (d - \ell^2 \Delta d).
\end{equation}

\begin{Remark}

The structure of our theory is similar in many respects to phase field models of fracture based on the top-down, critical energy release rate approach for polymers \citep[cf., e.g.,][]{Miehe14SchanzelJMPS,wu2016stochastic} and \citet{raina2016phase}. A reader familiar with these works might expect to see the degradation function $g(\dee)$ multiplying the entire undamaged free energy
\begin{equation}
    g(\dee) \underbrace{\left(\varepsilon_\mat^0-\thet\eta_\mat\right)}_{\psi_\mat^0},
    \label{free_energy_wrong}
\end{equation}
instead of degrading only the internal energy as in \eqref{network_internal_energy}. This choice would induce a \emph{free energy of scission} as a material parameter instead of the internal energy of scission, $\varepsilon_\mat^f$,  appearing in our theory. The advantage of our proposed form is that it respects the energetics of molecular bond dissociation.

\end{Remark}

\section{Summary of the governing partial differential equations for the specialized theory}
 \label{summspecialzed}

\begin{enumerate}
\item
\textbf{Balance of linear momentum}:    

\begin{equation}
 \Div \T_\mat  +\bfb_{0\mat}=\rho_\mat \ddot\y,
 \label{macfb2final}
\end{equation}
with $\T_\mat$
    given by
 \begin{equation}
\T_\mat   =   \bar\gee  \,
\left( J^{-2/3} \F - \bar{\lambda}^2 \F^{-\trans} \right) 
+  (1-\dee)^2 K(J-1)J \F^{-\trans},
 \label{summarystress}
\end{equation}
where
\begin{equation}
\bar\gee  \Def (N\,k_B\,\thet)
\left(\dfrac{1 }{3} \dfrac{\sqrt{n}}{\lambdabar\lambda_b}\right)\,
\calL^\inv\left(\dfrac{\bar{\lambda} \lambda_b^\inv}{\sqrt{n}}\right), 
 \label{summarystress1}
\end{equation}
is a generalized shear modulus in which  the effective stretch is $\lambdabar =\sqrt{\Cbar/3}$, and the  effective bond stretch $\lambda_b$ is determined by solving the implicit equation
 \begin{equation}
  (1-d)^2 E_b \left(\lambda_{\text{b}}-1\right)  -       k_B\thet\left(\dfrac{\lambdabar}{\sqrt{n}\lambdab^2}\right)\calL^\inv\left(\dfrac{\bar{\lambda} \lambdab^\inv}{\sqrt{n}}\right)=0.
    \label{summary_condition}
\end{equation}
Here, $\mathcal{L}^{-1}$ is the inverse of the Langevin function $\mathcal{L}(x) = \coth x - x^{-1}$.

\item \textbf{Microforce balance. Non-local evolution  equation for $\dee$}:


 \begin{equation}
\begin{split}
 \zeta \deedot  &    = 2(1-\dee)\calH   -\varepsilon_\mat^f(\dee-\ell^2\Delta\dee),
\end{split}
\label{evophifinal}
\end{equation}
with 
\begin{equation}
\varepsilon_\mat^f = N\,n\,\varepsilon_b^f,
\label{evophifinal1}
\end{equation}
a fracture energy, and
a history field function $\calH$ defined by
\begin{equation}
    \mathcal{H}(t) \Def \max_{s \in [0,t]} 
    \left\langle \hat{\varepsilon}_\mat^0 (\lambda_b(s), J(s)) - \varepsilon_\mat^f/2
    \right\rangle,
\end{equation}

where at each $s\in[0,t]$,
\begin{equation}
 \hat{\varepsilon}_\mat^0 (\lambda_b(s), J(s))  =
  \dfrac{1}{2} N\,n\, E_b \left(\lambda_b(s)-1\right)^2 +  \half K (J(s)-1)^2  .
 \label{evophifinal3}
\end{equation}
\end{enumerate}
 
The  theory  involves  the following  material parameters:
\begin{equation}
 N, \qquad n, \qquad E_b,\qquad K,   \qquad    \varepsilon_b^f, \qquad  \ell,  \qquad \text{and} \qquad  \zeta.
\label{evofinal4}
\end{equation}
Here, 
$N$  is the number of chains per unit volume; $n$ is the number of Kuhn segments in a chain; $E_b$ represents the modulus related to stretching of the bonds (Kuhn segments) of the polymer molecules; $K$ represents the bulk modulus of the material;  
$\varepsilon_b^f$, a bond dissociation energy per unit volume;    
$\ell$ is  a characteristic length scale of the gradient theory under consideration; 
and $\zeta$ is a kinetic modulus for the evolution of the damage. All parameters are required to be positive.
%


The boundary conditions for these partial differential equations  have been discussed previously   in Section~\ref{boundaryconds}.

\section{Application of the model to flaw sensitivity in elastomers}
\label{application}

\begin{figure}[htbp]
    \centering
    \includegraphics[scale=1]{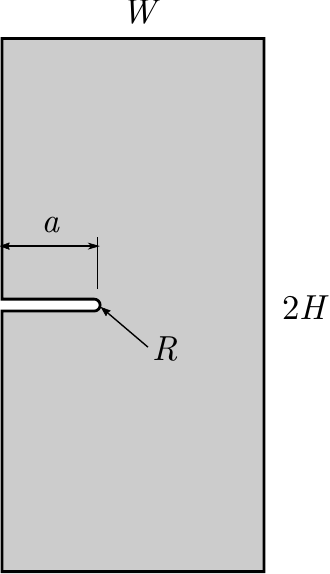}
    \caption{Geometry of single edge notch specimen used for flaw size sensitivity study. We take $W = 10a$, $H=20a$}
    \label{fig:sent_geometry}
\end{figure}

We demonstrate the behavior of the model through an example. We apply the phase field model to study the sensitivity of rupture to flaw size in an elastomeric material. In particular, we aim to show how the bond deformation local to the crack tip leads to behavior that is quantitatively different than that predicted by the Griffith theory at small length scales. 
We consider the problem of plane stress, mode I loading of a series of bodies containing a single edge notch (see Figure \ref{fig:sent_geometry}). The proposed model is implemented in the commercial finite element code Abaqus \citep{abaqus} through user-defined elements. A brief description of the implementation is given in an appendix.

Referring to Figure \ref{fig:sent_geometry}, each specimen is defined by the in-plane width $W$, half-height $H$, notch depth $a$, and notch root radius $R$. The considered notch lengths $a$, normalized by the material parameter $\ell$, are
\begin{equation}
    a / \ell = \lbrace \text{0.16, 0.33, 0.63, 1.90, 6.35, 10.79, 63.49} \rbrace.
\end{equation}
  For each notch length  we scale the specimen so that the relative crack depth and reamining ligament are identical, with the notch being shallow with respect to the ligament. We take $W = 10a$ and $H=20a$. The main interest is in notches that are sharp with respect to the chain network size, so we keep the notch radius fixed at the value $R=0.063 \ell$ for all cases. Each specimen is stretched vertically until the specimen ruptures. The chosen material properties are given in Table \ref{table:props}.
The nominal stretch rate $\dot{\lambda}$ and the phase field kinetic parameter $\zeta$ are selected such that rate effects are negligible.

\begin{table}[htbp]
    \centering
    \caption{\label{table:props} Material properties for the mode I simulations.}
    \begin{tabular}{ccccc}
        \toprule
        $n$ & $E_b / k_B \vartheta$ & $K / N k_B \vartheta$ & $\varepsilon_b^f / k_B \vartheta$ & $  \zeta / (N k_B \vartheta /\dot{\lambda})$ 
        \\
        \midrule
        3 & 1000 & 5000 & 100 & 0.045
        \\
        \bottomrule
    \end{tabular}
\end{table}

\begin{figure}[htbp]
    \centering
    \begin{subfigure}{1\textwidth}
        \includegraphics[width=0.9\textwidth]{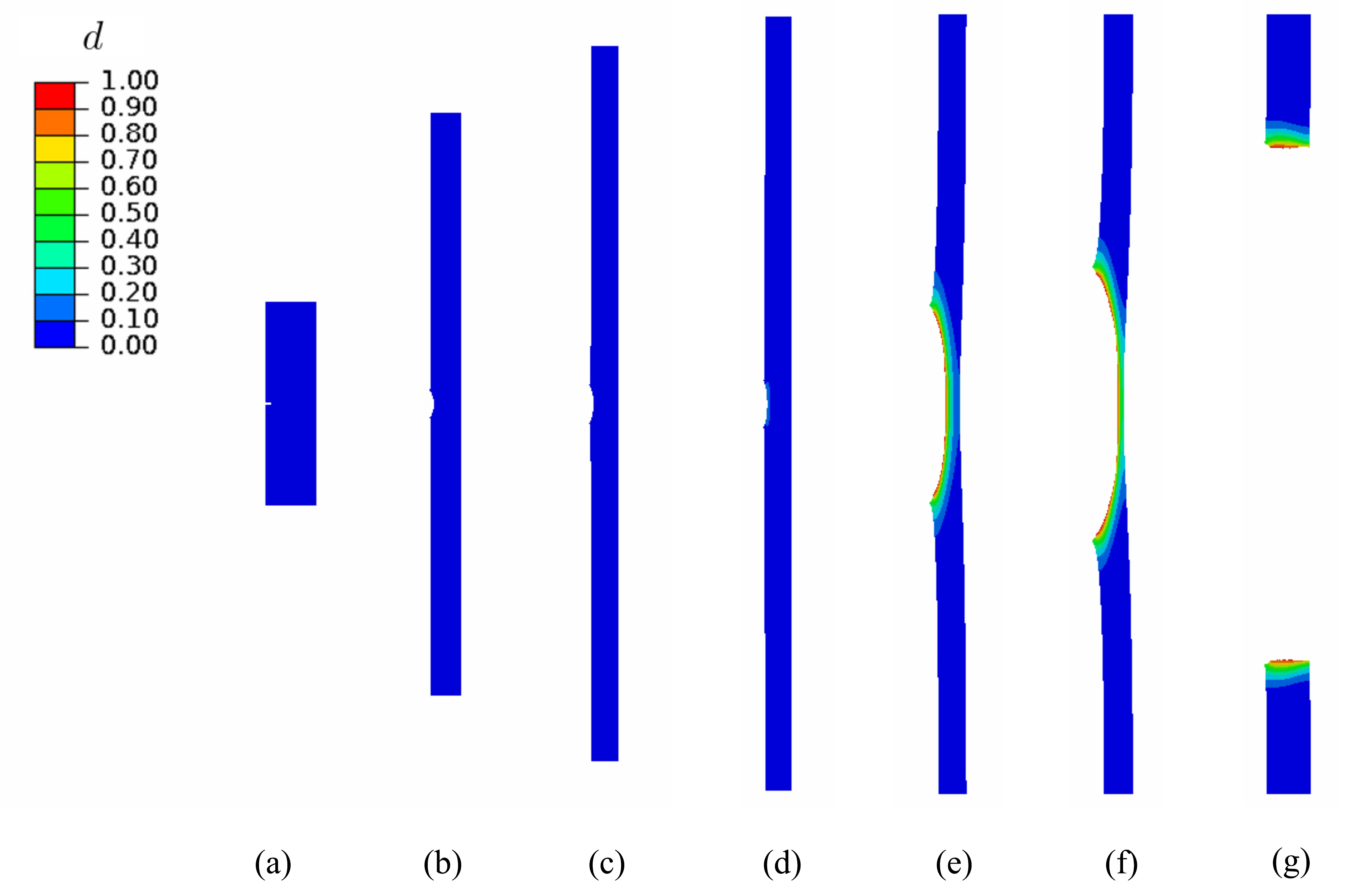}
    \end{subfigure}
    \\
    \begin{subfigure}{0.7\textwidth}
        \includegraphics[width=1.0\textwidth]{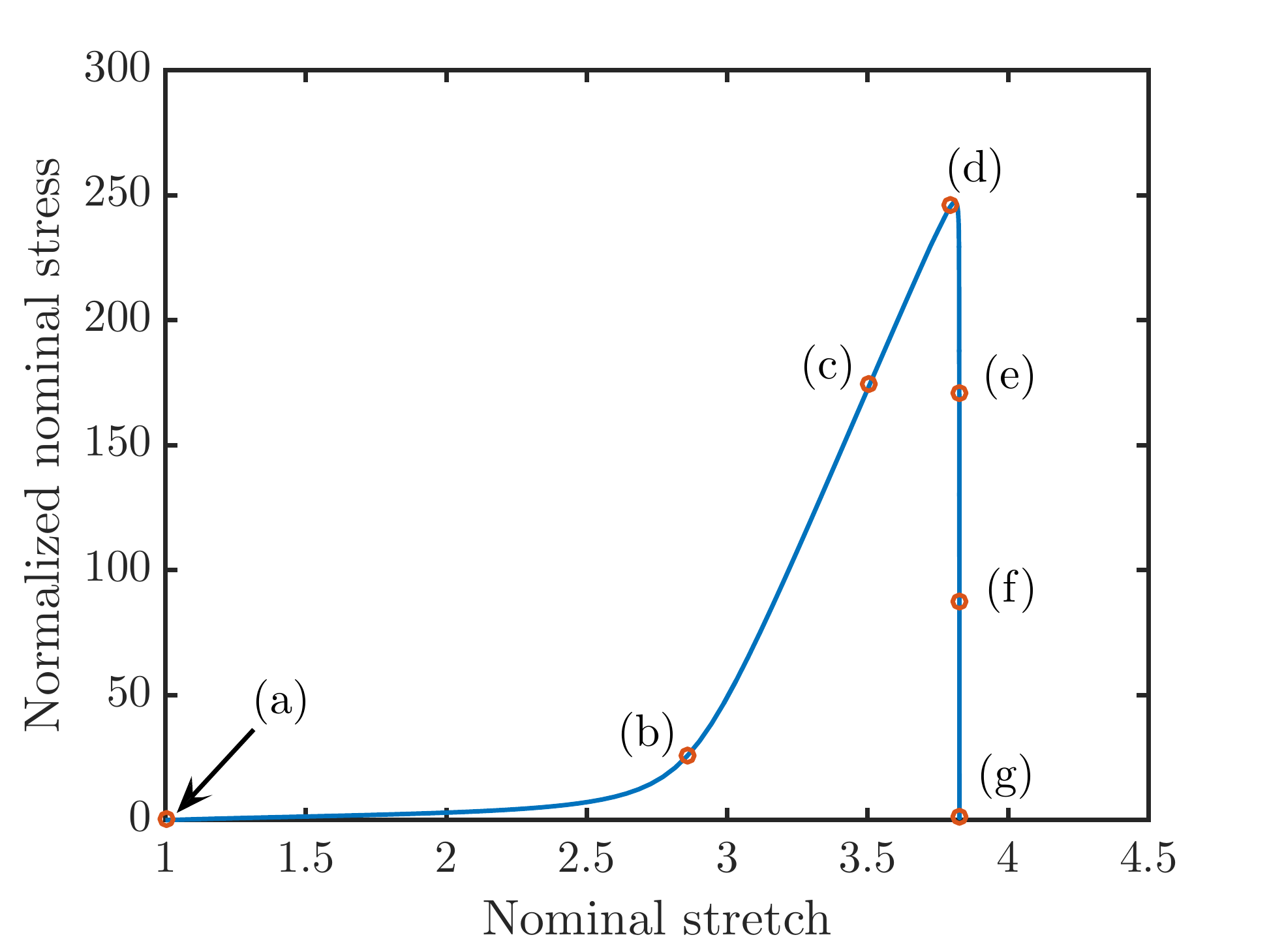}
    \end{subfigure}
    \caption{Progressive damage and rupture for the case $a / \ell = 1/3$. Top: Images of the progressive stretch, damage, and rupture in the specimen, with contours of the phase-field $d$. To aid visualization of the damage, elements with an average value of $d > 0.95$ are removed from the plot. Bottom: Computed nominal stress vs.\ nominal stretch. The points a--f correspond to the images above.}
    \label{mode_I_progressive_failure}
\end{figure}

Snapshots from the case $a / \ell = 0.33$ are shown in Figure \ref{mode_I_progressive_failure} to illustrate the character of the model. The specimen first stretches elastically, showing non-Gaussian stiffening typical of the Arruda-Boyce model (point (b)). When the internal energy reaches the critical point at the notch (point (d)), chain scission damage commences and propagates across the ligament as a crack-like feature (points (d)-(f)). (Highly damaged elements are removed from the snapshots in Figure \ref{mode_I_progressive_failure} to aid visualization of the crack growth).
Eventually, the ligament fails, separating the specimen into two unloaded parts (point (g)).

\begin{figure}[htbp]
    \centering
    \includegraphics[width=0.7\textwidth]{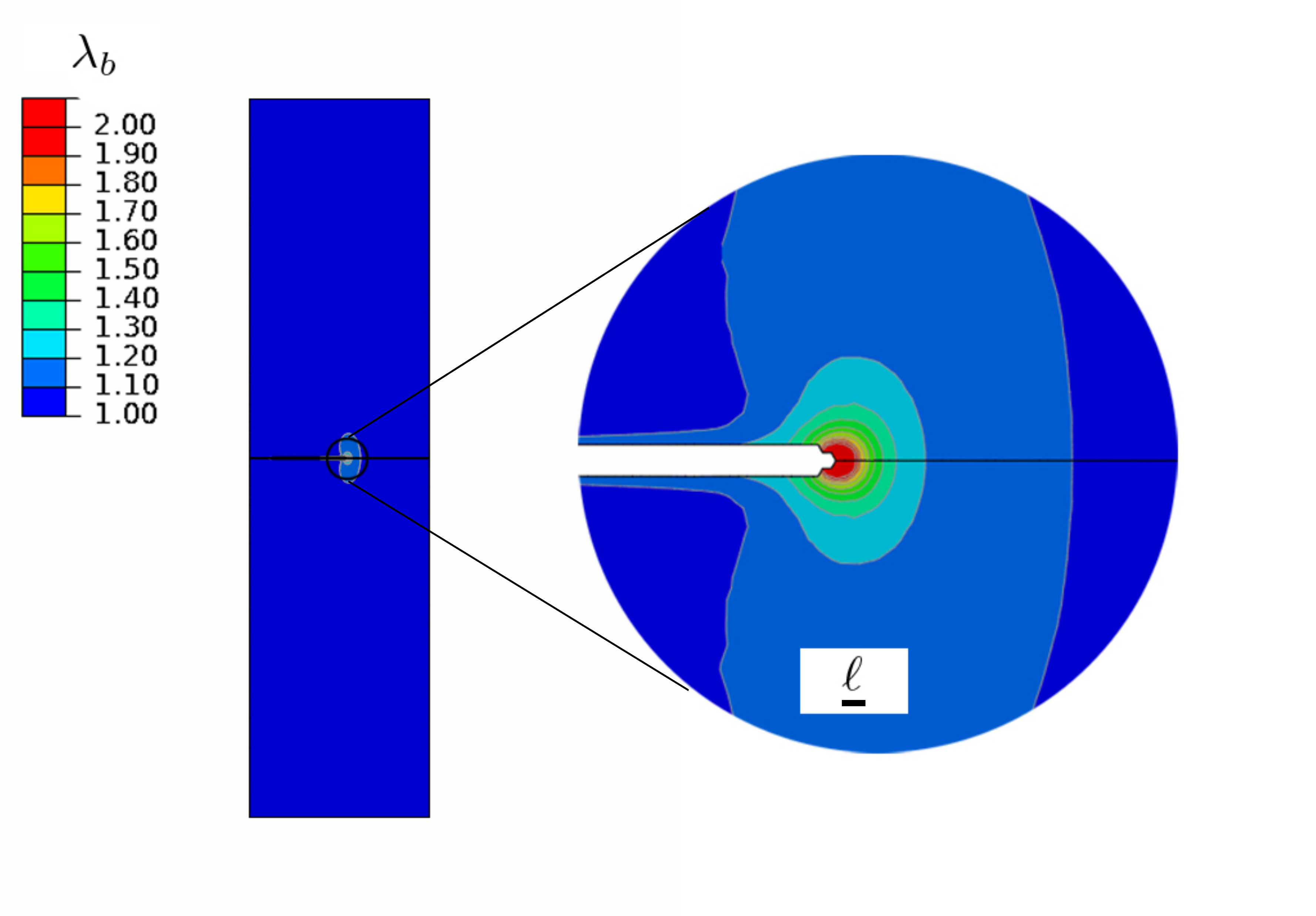}
    \caption{Contours of bond stretch $\lambdab$ during the fracture process for a large notch, $a / \ell = 10.79$. The bond deformation stretch is appreciable only in a small zone near the crack tip. (Contours are plotted on the reference configuration. Elements are removed from the visualization when $d > 0.95$ at all element nodes.)}
    \label{fig:notch_anim_frame_large}
\end{figure}

\begin{figure}[htbp]
    \centering
    \includegraphics[width=0.75\textwidth]{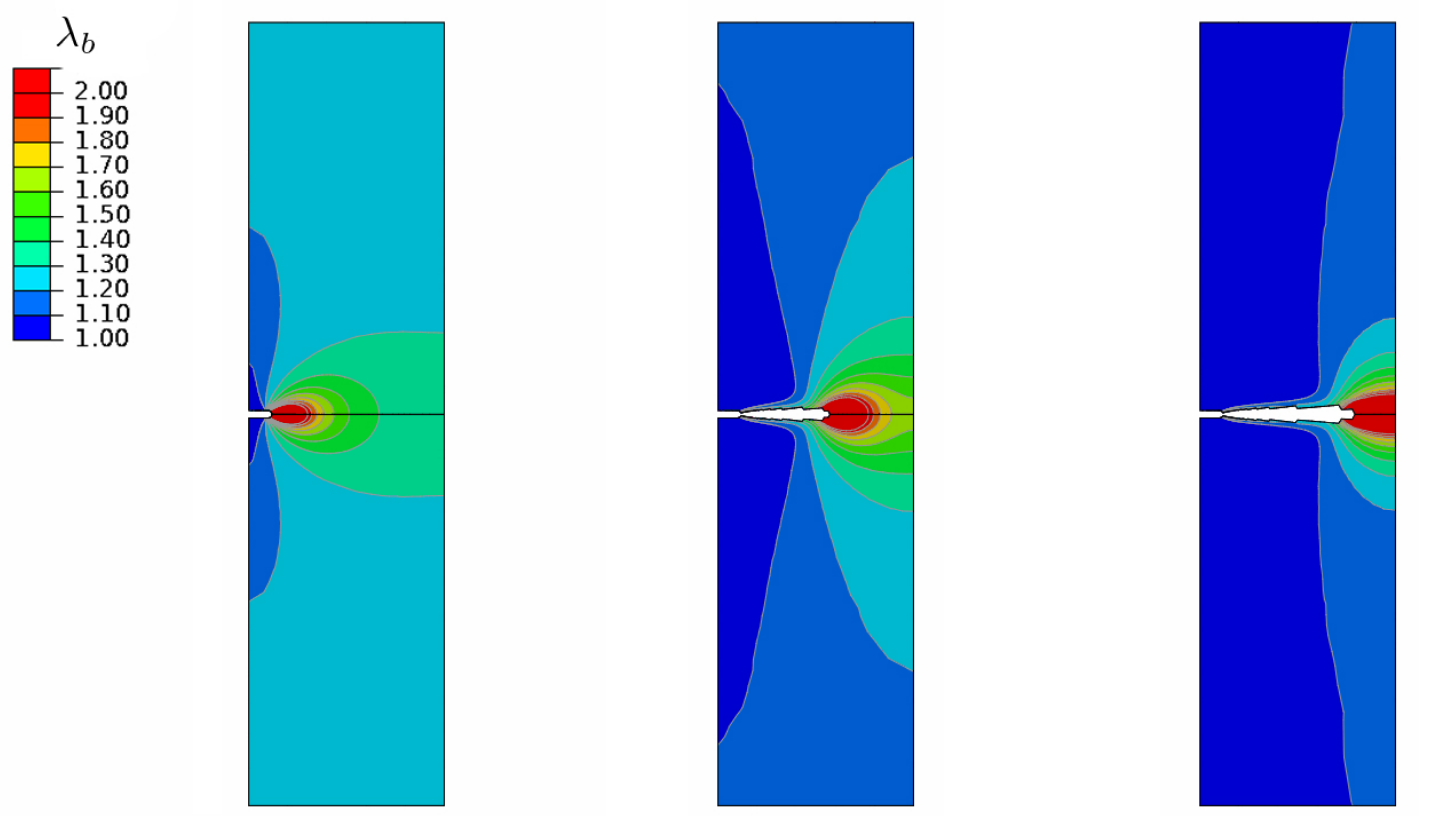}
    \caption{Contours of bond stretch $\lambdab$ during the fracture process for a small notch, $a / \ell = 1/3$. For a small flaw, significant bond deformation occurs throughout the entire specimen. Snapshots are shown at three different stages of crack propagation. (Contours are plotted on the reference configuration. Elements are removed from the visualization when $d > 0.95$ at all element nodes.) }
    \label{fig:notch_anim_frame_small}
\end{figure}

We now examine the role of the bond mechanics on the overall response. In Figures  \ref{fig:notch_anim_frame_large} and \ref{fig:notch_anim_frame_small}, we plot contours of the bond deformation stretch $\lambdab$ during the deformation process. Highly damaged elements ($d > 0.95$) are again hidden from view. The contours are plotted on the reference configuration to highlight the extent of crack propagation relative to the initial specimen geometry.
A  specimen  with a  large flaw, $a / \ell = 10.79$, is shown in Figure \ref{fig:notch_anim_frame_large}. The crack has propagated halfway through the specimen. In this case, the bond stretching is limited to a small region in the vicinity of the crack tip on a scale comparable to $\ell$; the majority of the material displays negligible bond deformation and is thus well described by the Arruda-Boyce model. This plot illustrates that for large flaws (with respect to $\ell$), the mechanics of the crack tip and of the body as a whole are separated in scale, and top-down fracture mechanics may be applied.

Figure \ref{fig:notch_anim_frame_small} depicts the bond deformation for a small flaw ($a / \ell = 0.33$). The image in the first frame is taken when extension of the notch is imminent. The bond deformation stretch is appreciable throughout the entire specimen, and the overall response is strongly influenced by the mechanics of the bonds. 
This Figure illustrates why the small scale behavior of the system is a useful limit to model, as the stress-bearing capacity of the material is being used efficiently. At the point of rupture, most of the specimen is being stretched close to the theoretical limit set by the binding energy of the bonds in the  chain.

\begin{figure}[htbp]
    \centering
    \includegraphics[width=0.7\textwidth]{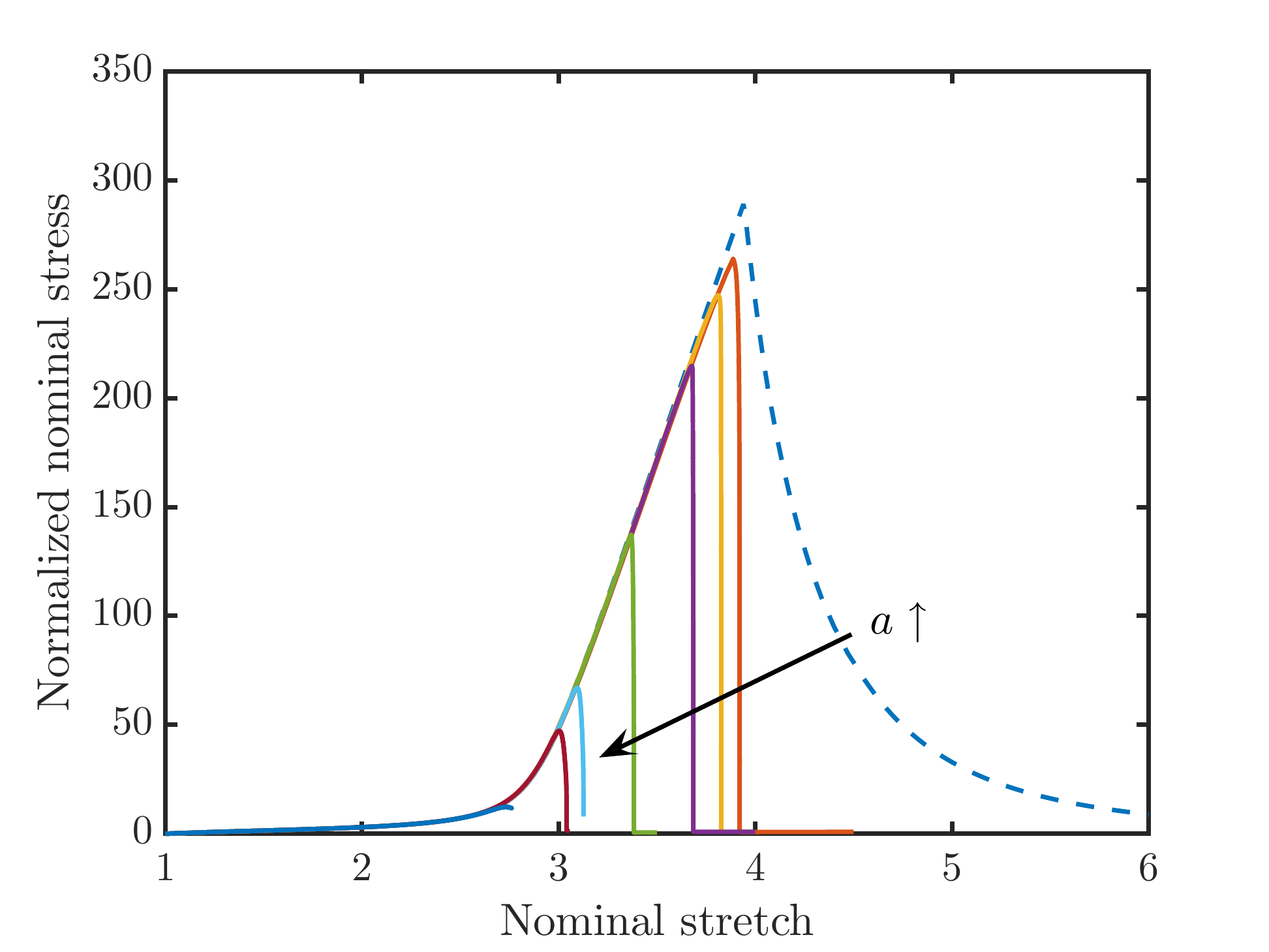}
    \caption{Nominal stress (normalized by $N k_B \vartheta$) vs.\ nominal stretch for all single edge notch tests, with $a / \ell = \lbrace 0.16, 0.33, 0.63, 1.9, 6.4, 10.8, 63.49 \rbrace$ (solid lines). The dashed curve shows the behavior of material with no defects, computed from a single material point.}
    \label{force_displacement_all}
\end{figure}

The nominal stress  versus nominal stretch response for all considered cases is shown in Figure \ref{force_displacement_all}. 
These results may be considered representative of the macroscopic behavior that would be measured on a sample of material that contains a flaw of a certain size.
For comparison, the ideal behavior of the material with no flaws is included on the plot as a dashed curve.\footnote{
The ideal behavior is computed by applying a uniaxial, plane stress deformation to a single material point, neglecting the nonlocal $\nabla d$ term.}  
The nominal stretch and stress to rupture decrease as the flaw size increases, with the nominal stress attained by the smallest samples approaching the ideal strength of the material.  

\begin{figure}[htbp]
    \centering
    \begin{subfigure}{0.49\textwidth}
        \includegraphics[width=1.0\textwidth]{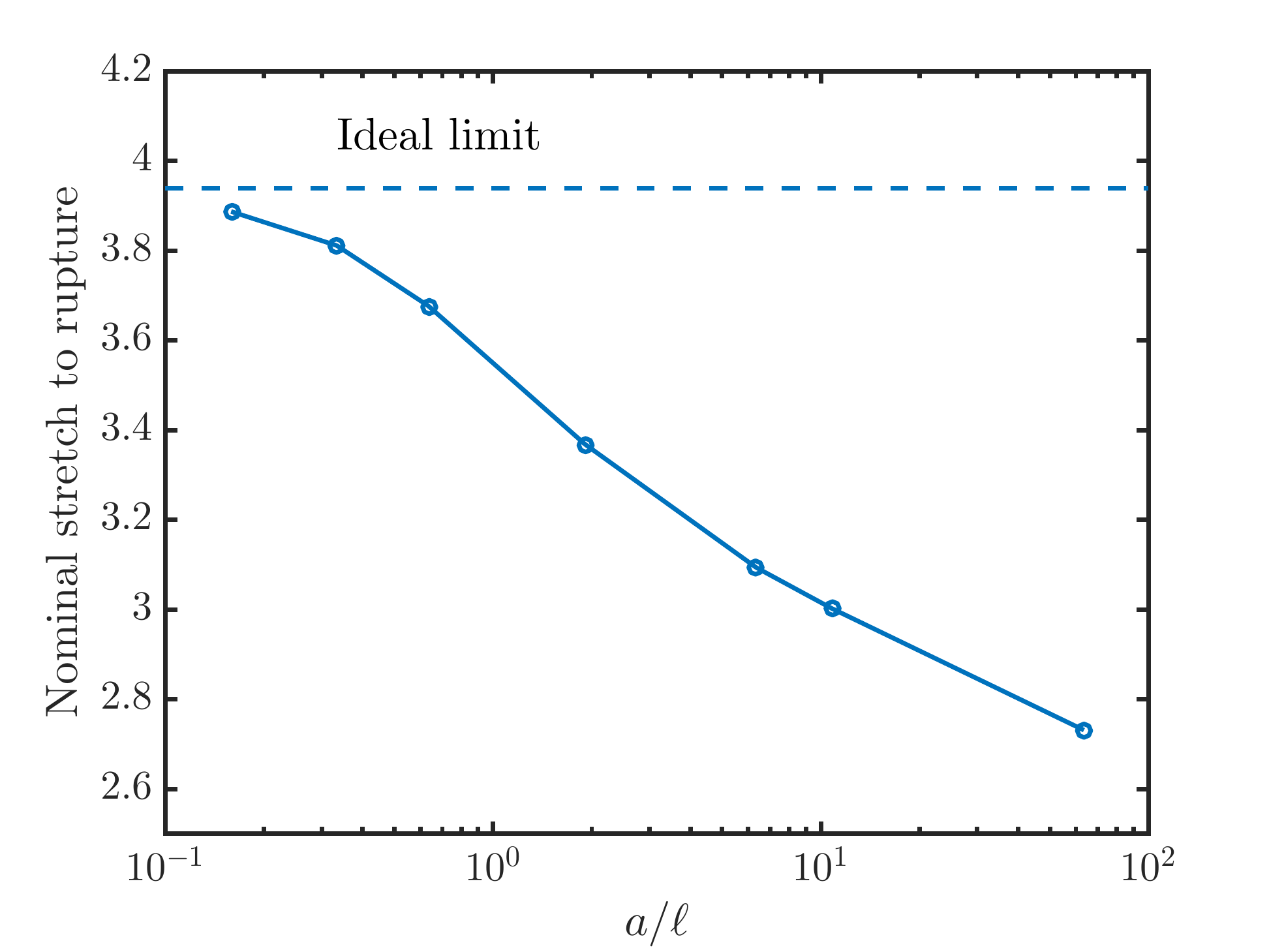}
        \caption{ }
    \end{subfigure}
    \begin{subfigure}{0.49\textwidth}
        \includegraphics[width=1.0\textwidth]{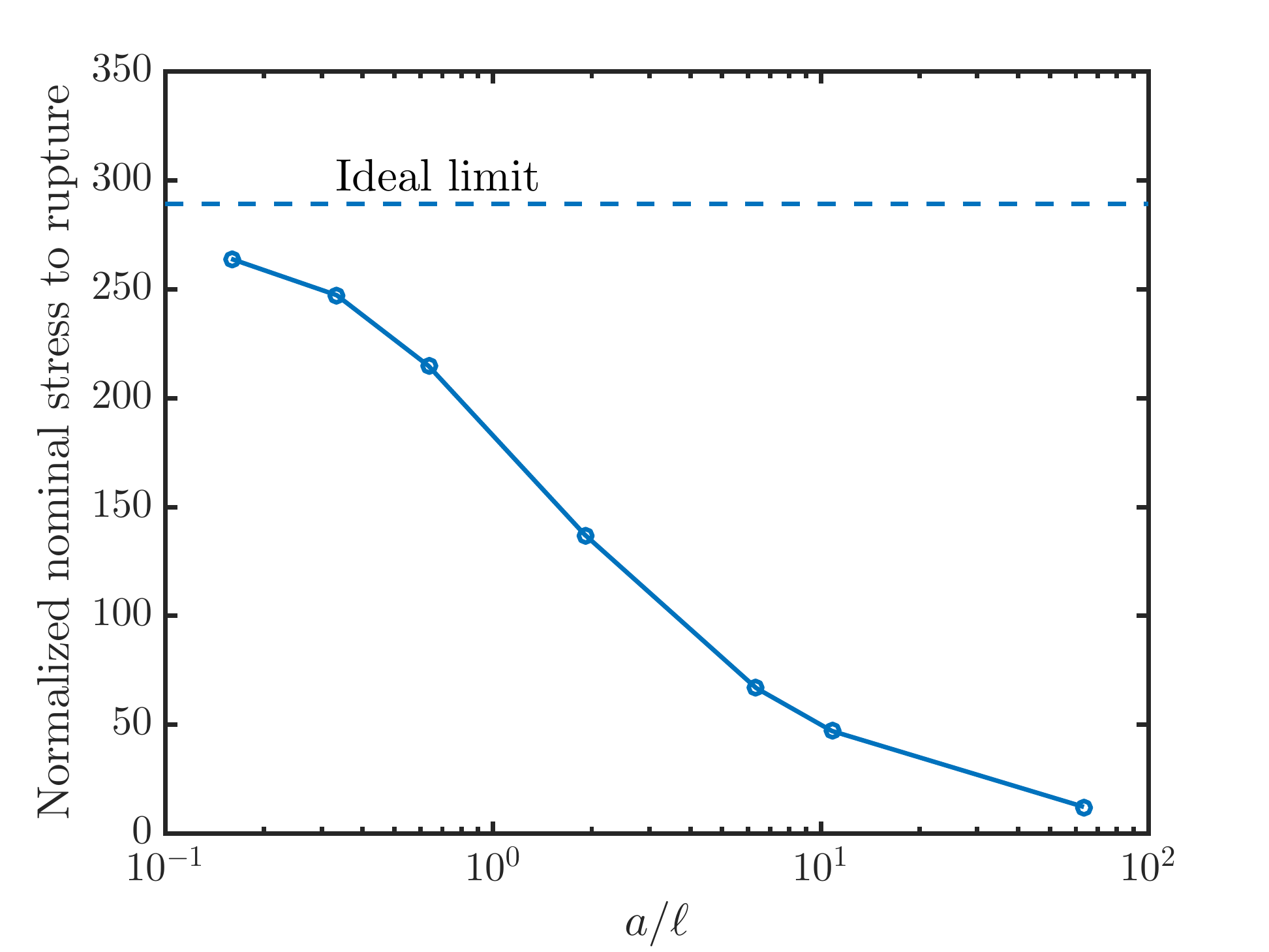}
        \caption{ }
    \end{subfigure}
    \caption{Behavior in the single edge-notch test as a function of $a / \ell$. (a) Nominal stretch to rupture vs.\ normalized crack size. (b) Nominal stress to rupture (normalized by shear modulus) vs.\ normalized crack size. The stretch and stress to rupture increase with decreasing flaw size, but are capped by the ideal limit of the material. The sensitivity to flaw size decreases as the ideal limit is approached.}
    \label{flaw_sensitivity_plots}
\end{figure}

The predictions of the flaw sensitivity are summarized in Figure \ref{flaw_sensitivity_plots}, in which we plot the nominal stretch and stress to rupture against the normalized flaw size. Rupture is defined here as the point where the load maximum is reached. Similar behavior is seen in both the stretch and stress metrics.
The strength of the specimen is strongly dependent on the flaw size when the flaw is large with respect to $\ell$. For flaws comparable in size to $\ell$ or smaller, the stress concentrating effect of the flaw is weak, and the body is limited globally by the ideal load bearing capacity of the bonds in the polymer chains. The rupture stretch and stress show a weak dependence on flaw size in this regime, which is a manifestation of the widespread influence of the bond deformation mechanics. This flaw-tolerant behavior can be deliberately engineered in composites and synthetic bio-inspired materials if the length scales of the soft phases can be suitably controlled.

\subsection{Connection with the critical energy release rate}
\label{energy_release_rate}

We now compare the proposed model of rupture to the classical critical energy release rate condition for crack propagation. 
We use the same geometry and loading as in the previous section. 
As there is no sharply defined point for the initiation of crack propagation in the phase field model, we have chosen to define a crack propagation condition based on the phase field attaining a critical value a short distance ahead of the notch root. 
We define crack propagation to occur when the phase field attains the condition $d = 0.95$ at a distance of $4R$ ahead of the initial notch. 
For each single edge notch specimen, we determine the nominal stretch when this propagation condition is met. The resulting stretch values are recorded in Table \ref{tab:crack_prop_stretch}.

For comparison, we performed additional numerical simulations to compute the critical energy release rate at the point of crack propagation. These calculations employ the built-in Abaqus implementation of the virtual crack extension method to compute the J-integral \citep{parks:1977}. To simplify the calculations, only the elastic response is modeled, and the coupling to the phase field is suppressed.\footnote{We implement the elastic material behavior through the UMAT user-defined material interface in Abaqus;  \citep[see][]{Mao17TalaminiEML}.}
Each specimen geomtery is stretched to the appropriate value in Table \ref{tab:crack_prop_stretch}, and the energy release rate computed.

\begin{table}
    \centering
    \caption{Stretch to crack propagation from phase field model \label{tab:crack_prop_stretch}}
    \begin{tabular}{cc}
    \toprule
    Flaw size $a / \ell$ & Nominal stretch at crack propagation
    \\
    \midrule
    0.16 & 3.92
    \\
    0.33 & 3.83
    \\
    0.63 & 3.68
    \\
    1.90 & 3.37
    \\
    6.35 & 3.08
    \\
    10.79 & 2.97
    \\
    63.49 & 2.44
    \\
    \bottomrule
    \end{tabular}
\end{table}

\begin{figure}[htbp]
    \centering
    \includegraphics[width=0.6\textwidth]{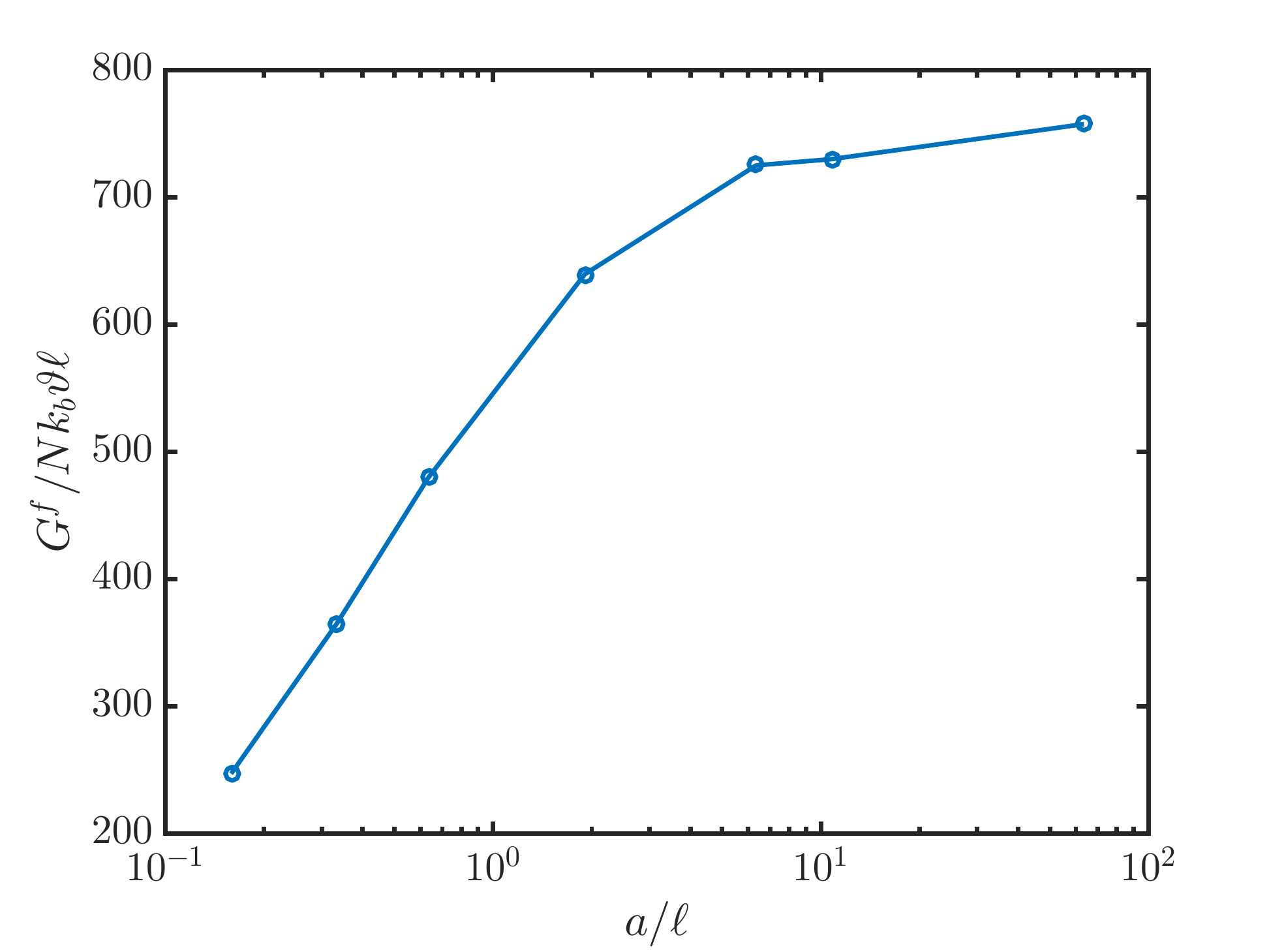}
    \caption{Energy release rate at the point of crack propagation as a function of $a / \ell$. As $a / \ell$ increases, the energy release rate at the point of crack propagation approaches a constant value, showing that the model is consistent with the classical Griffith approach. }
    \label{fig:energy_release_rate}
\end{figure}

The computed energy release rates at the point of crack propagation (denoted by $G_f$)  are plotted in Figure \ref{fig:energy_release_rate} against the flaw size. 
The results are presented in dimensionless form.
As the considered flaw size increases, the energy release rate asymptotically approaches a constant value, which represents the macroscopic critical energy release rate for the material. This reinforces the tenet that large systems can be characterized by the top-down approach of fracture mechanics, with the details of the crack tip behavior lumped into a single parameter. Conversely, the energy release rate for the propagation of small flaws--- say $a / \ell < 10$ here ---is not a material parameter, rather, it depends on the geometry and the material behavior in the crack tip vicinity. The proposed model bridges the scales from where the bond deformation mechanics matter, to the macro-scale.

\section{Conclusions}
\label{conclusions}

We have formulated a model for progressive damage and rupture in elastomeric materials which accounts for the underlying microscopic behavior of molecular bond deformation and scission. Adopting this microscopic view should be useful for modeling elastomeric materials at small  length scales, such as occurs in nano-composites and bio-inspired composites. When the length scales are very small, the flaw sensitivity diverges from the predictions of classical energy release rate-based fracture mechanics, with the material showing much lower sensitivity to flaws than predicted by classical fracture mechanics. The proposed model represents a step towards capturing this behavior and optimizing such materials at the small scale.

The model describes damage initiation, propagation, and full rupture in polymeric materials, and detects the transition from flaw-insensitive behavior at small scales to flaw-sensitive propagation of sharp cracks at large length scales.

The present work is dedicated to the simplest case of elastomeric materials, which can undergo large reversible deformations with negligible rate-dependent dissipation. It would be useful to extend these ideas to materials which exhibit additional dissipation mechanism (e.g., viscoelasticity  and Mullins effect) that  accompany   the rupture process, which can be exploited as toughening mechanisms \citep{Zhao14SoftMatt,Ducrot14ChenScience,Mao17LinJMPS}.

\section*{Acknowledgements}

Support from Exxon-Mobil Research through the MIT Energy Initiative is gratefully acknowledged.

\clearpage

\appendix
 
\section{Detailed derivation of the theory}
 \label{appendix1}

In this Appendix we give details of our phase-field theory for fracture of a  finitely-deforming elastic solid.

\subsection{Kinematics}
\label{kinematics}

Consider  a  macroscopically homogeneous body $\text{B}$ with the region of space it occupies in a fixed  reference configuration, and denote by $\bfX$ an arbitrary material point of $\text{B}$.
A  motion  of
$\text{B}$ is then a smooth one-to-one mapping
\begin{equation}
\bfx =
\y(\bfX,t),
\label{kin0}
\end{equation}
 with   deformation gradient,  velocity, and acceleration  given by
%
\begin{equation}
\bfF = \nabla \y, \qquad \bfv = \dot{\y}, \qquad \dot\bfv = \ddot{\y}.
\label{kin1}
\end{equation}
As is standard, we assume that
\begin{equation}
J \Def \Det \,\bfF >0.
\label{kin3}
\end{equation}
%
The right and left
Cauchy-Green  tensors are  given respectively by
\begin{equation}
\C=\F{}^{\trans}\,\F,
 \qquad
\bfB=\F\vvs\F{}^{\trans}.
\label{kin10}
\end{equation}

We denote  the distortional  (or volume preserving)  part of $\F$
 by
\begin{equation}
\Fbar \Def J^{ \,-1/3}\,\F,\quad \det
\Fbar=1,
\label{distort1a}
\end{equation}
 and correspondingly let
\begin{equation}
\Cbar \Def  \Fbar^{\trans}\Fbar=J^{-2/3}\C, \qquad \Bbar \Def\Fbar \Fbar^{\trans} =J^{-2/3}\bfB
\label{fe3a}
  \end{equation}
denote the distortional   right  and left Cauchy-Green tensors.

%

We denote    an 
 arbitrary {part}   of  $\B$  by   $\p$, and by  $\bfn_\mat$ the outward unit normal on the
boundary $\partial\p$ of $\p$. 
Further, henceforth  we   use a subscript ``$\mat$''  to refer to quantities measured in the reference configuration.

\subsection{Effective bond stretch}

Following \cite{Mao17TalaminiEML} we  introduce a dimensionless  positive-valued  internal variable, 
$$\lambda_b>0,$$
to represent (at the continuum scale) a measure of the stretch  of the Kuhn segments of the polymer chains. 
We call $\lambda_b$ the \emph{effective bond stretch}.

\subsection{Phase-field} 
To describe fracture  we introduce  an   \emph{order-parameter} or \emph{phase-field},
\begin{equation}
\text{$\dee(\X,t)\in [0,1]$}.
\label{varphidef}
\end{equation}
  If $\dee=0$ at a point then that point is  intact, while  if $\dee=1$ at some point, then that point is fractured. Values of $\dee$ between zero and one correspond to partially-fractured material.  
We assume that $\dee$  grows  montonically so that 
\begin{equation}
 \deedot(\X,t) \ge 0,\label{dissipativephi}
\end{equation}
which is a  constraint  that  represents the usual  assumption that microstructural changes leading to  fracture are \emph{irreversible}.

%
%
%
%
%
%

\subsection{Principle of virtual power. Macroscopic and microscopic force balances}
 \label{vpower}

We follow    \cite{gurtin1996,gurtin2002} and \cite{gurtin-fried-anand} to derive macroscopic and microscopic force balances derived via the principle of virtual power.
In developing our theory we take the ``rate-like'' kinematical descriptors to be
   $\dot\y$,
$\Fdot$,  $\lambdabdot$, and  $\deedot$,      and  also the gradient  $\nabla\deedot$.
%
In exploiting the principle of virtual power we note that  the rates   $\dot\y$ and 
$\Fdot$   are  not independent ---
they are constrained by  (cf. \eqref{kin1})
\begin{equation}
 \nabla \dot\y  =\Fdot.
 \label{vpower0}
\end{equation}

With each evolution of the body we associate macroscopic and
microscopic force systems.   The macroscopic system is defined by:
\begin{itemize}
\item[(i)]
a {traction} $\bft_\mat(\bfn_\mat)$ (for each unit vector $\bfn_\mat$) that
expends power over the velocity $\dot\y$;
\item[(ii)]  a generalized  external   body $\bfb_\mat$ that  expends
power over $\dot\y$, where
\begin{equation}
\bfb_\mat=\bfb_{0\mat}-\rho_\mat\ddot\y,
\label{inertial}
\end{equation}
 with
  $\bfb_{0\mat}$ the non-inertial body force and $\rho_\mat$ the mass
density in the reference configuration; and 
\item[(iii)]

 a stress $\T_\mat$ that
expends power over the  distortion rate $\Fdot$. 
\end{itemize}
The
microscopic system, which is non-standard, is defined by:
\begin{itemize}
\item [(a)]  a   {scalar microscopic
    force} $f$  that expends power over the 
    rate $\lambdabdot$;  %
\item [(b)]  a   {scalar microscopic
    stress} $\varpi$  that expends power over the 
    rate $\deedot$;  %
      \item [(c)] a   vector  microscopic
    stress    $\bfxi$ that expends power over the   gradient  
    $\nabla\deedot$;  and
     \item [(d)] a scalar
     {microscopic traction} $\zeta(\bfn_\mat)$ that expends
    power over $\deedot$.
    %
%
\end{itemize}

 We characterize the force systems
through the manner in which these forces expend power. That is,
given any part $\p$, through the specification of
$\calW_{\text{ext}}(\p)$, the power expended on $\p$ by material
  external to $\p$, and $\calW_{\text{int}}(\p)$, a
concomitant expenditure of power   within $\p$. Specifically,
\begin{equation}
\left.
\begin{aligned}
\calW_{\text{ext}}(\p) & =
\int\limits_{\partial\p}\!\bft_\mat(\bfn_\mat) \cdot\dot\y \,da_\mat +
\int\limits_{\p}\!\bfb_\mat\cdot\dot\y\,dv_\mat 
+\int\limits_{\partial\p}\!\zeta(\bfn_\mat)\vs\deedot\,da_\mat,
\\[4pt]
\calW_{\text{int}}(\p) & =
\int\limits\limits_{\p}\!\Big(\T_\mat\tendot\Fdot +f\lambdabdot+\varpi \deedot+
\bfxi \!\cdot\nabla\deedot
  \Big)
  \,dv_\mat,
\end{aligned}
\right\}
 \label{vpower1}
\end{equation}
where, $\T_\mat$, $f$, $\varpi$,     and
$\bfxi$, 
 are defined over the body for all
time.

Assume that, at some arbitrarily chosen but {\it fixed time\/},
the fields $\y$,  $\F$, $\lambda_b$,  and $\dee$ are known, and consider the fields $\dot\y$, $\Fdot$, $\lambdabdot$,  and 
$\deedot$    as virtual velocities to be specified independently in a
manner consistent with \eqref{vpower0}; that is, denoting the
virtual fields by $\tilde{\y}$, $\Ftilde$,  $\lambdabtilde$,  and $\deetilde$ to differentiate them from fields associated with the actual
evolution of the body, we require that
\begin{equation}
 \nabla \tilde\y  =\Ftilde .
\label{vpower2}
\end{equation}
Further,  
we
define a  {generalized virtual velocity} to be a list
$$
\calV = (\tilde{\y},\Ftilde,\lambdabtilde,\deetilde),
$$
consistent with \eqref{vpower2}.
%


%
We refer to a  macroscopic virtual field $\calV$  as \emph{rigid}  
  if it satisfies 
\begin{equation}
(\nabla\tilde{\y} ) =\tilde\F =  \bfOmega\F  \qquad \text{together with}  \qquad  \lambdabtilde=0, \qquad  \deetilde=0,
\label{rigid1a}
\end{equation}
with $\bfOmega$ a spatially constant skew tensor.

%
Next, writing
\begin{equation}
\left.
\begin{aligned}
\calW_{\text{ext}}(\p,\calV) &=
 \int_{\partial\p}\bft_\mat(\bfn_\mat) \cdot\tilde{\y}\,da_\mat+
\int_{\p}\bfb_\mat\cdot\tilde{\y}\,dv_\mat
+
\int_{\partial\p}\zeta(\bfn_\mat)\,\deetilde\,da_\mat,\\
\calW_{\text{int}}(\p,\calV)
 &= \int_{\p}\!\Big(\T_\mat\tendot\Ftilde
+ f\lambdabtilde+\varpi\deetilde  + \bfxi \cdot \nabla\deetilde
   \Big)\,dv_\mat,
\end{aligned}
\right\}
\label{vpower3}
\end{equation}
respectively, for the external and internal expenditures of
 virtual power, 
the \emph{principle of virtual
power} consists of two basic requirements:
\begin{itemize}
\item [(V1)]  Given any part P,
\begin{equation}
\calW_{\text{ext}}(\prt,\calV) = \calW_{\text{int}}(\prt,\calV) \quad
\text{for all generalized virtual velocities } \calV.
\label{vpower4}
\end{equation}
\item [(V2)]    Given  any part $\p$ and a \emph{rigid}  virtual velocity $\calV$,
\begin{equation}
\calW_{\text{int}}(\p,\calV) =0 \text{ \  whenever $\calV$ is a rigid macroscopic virtual velociy.}
\label{vpower5}
\end{equation}
\end{itemize}
To deduce the consequences of the principle of virtual power, assume that \eqref{vpower4} and \eqref{vpower5} are  satisfied. Note that in applying the virtual
balance   we are at liberty to choose any $\calV$
consistent with the constraint  \eqref{vpower2}.

\subsection{Macroscopic force and moment balances}

Let $\lambdabtilde=0$ and    $\deetilde=0$. For
this choice of $\calV$, \eqref{vpower4} yields
\begin{equation}
\int_{\partial\p}\bft_\mat(\bfn_\mat) \cdot\tilde{\y}\,da_\mat +
\int_{\p}\bfb_\mat\cdot\tilde{\y}\,dv_\mat
= \int_{\p} \T_\mat \tendot\Ftilde\,dv_\mat= \int_{\p} \T_\mat \tendot\nabla\tilde{\y} \,dv_\mat,
\label{vpower9}
\end{equation}
which
may be rewritten as
\begin{equation}
\int_{\partial\p}\bft_\mat(\bfn_\mat) \cdot\tilde{\y}\,da_\mat   = \int_{\p}\Big( \T_\mat \tendot
\nabla\,\tilde{\y}-\bfb_\mat\cdot\tilde{\y}\Big)\,dv_\mat,
 \label{vpower11}
\end{equation}
and   using the divergence theorem  we may conclude that
$$
\int_{\partial\p}\bigl(\bft_\mat(\bfn_\mat)  -
\T_\mat\bfn_\mat\bigr)\cdot\tilde{\y}\,da_\mat + \int_{\p}(\Div\,\T_\mat +
\bfb_\mat)\cdot\tilde{\y}\,dv_\mat = 0.
$$
Since this relation must hold for all $\p$ and all $\tilde{\y}$,
standard variational arguments yield the   traction condition
\begin{equation}
\fbox{ \parbox{3cm} {\[  
\bft_\mat(\bfn_\mat) =\T_\mat\bfn_\mat\,,
\]}}  
  \label{vpower12}
\end{equation}
and the local macroscopic force balance
\begin{equation}
\fbox{ \parbox{4cm} {\[  
\Div\,\T_\mat + \bfb_\mat= \zed, 
\]}}  
  \label{vpower13}
\end{equation}
respectively.  

Next,  we deduce the consequences of   requirement (V2)  of the  principle of virtual power.
 Using \eqref{rigid1a} and \eqref{vpower3}$_2$,  requirement (V2)  of the  principle of virtual power leads to the requirement that
  \begin{equation}
 \int_{\p}\, (\T_\mat\F^\trans)\tendot \bfOmega\, dv_\mat=0.
 \label{rigid3}
 \end{equation}
Since $\p$ is arbitrary, we obtain that $(\T_\mat\F^\trans)\tendot \bfOmega=0$ for all skew tensors $\bfOmega$, which implies that $\T_\mat\F^\trans$ is \emph{symmetric}:
%
\begin{equation}
\fbox{ \parbox{3cm} {\[  
\T_\mat\bfF^{\trans} = \bfF\T_\mat^{\trans}. 
\]}}  
 \label{vpower14}
\end{equation}
\begin{itemize}
\item
\emph{In view of \eqref{vpower13} and \eqref{vpower14} the stress $\T_\mat$
represents the classical   {Piola stress}, with
\eqref{vpower13} and \eqref{vpower14}  representing the local
{macroscopic force and moment balances} in the reference
body}. 
\end{itemize}
Upon using  the expression \eqref{inertial} for  $\bfb_\mat$ in \eqref{vpower13}, we obtain the equation of motion
\begin{equation}
\fbox{ \parbox{4cm} {\[  
\Div\,\T_\mat + \bfb_{0\mat} = \rho_\mat\ddot\y.
\]}}  
  \label{vpower13b}
\end{equation}

\subsection{Microscopic force balances}

\begin{enumerate} 
\item Next, consider a generalized virtual velocity with
$\tilde{\y}={\bf 0}$ and $\deetilde=0$ and choose the virtual field $\lambdabtilde$
arbitrarily. %
The power balance \eqref{vpower4}  then yields
the   microscopic virtual-power relation
\begin{equation}
0
=
 \int\limits_{\p} 
f\lambdabtilde
  \,dv_\mat ,
\label{vpower26a}
\end{equation}
to be satisfied for all  $\lambdabtilde$ and all $\p$, and a standard argument yields the  microscopic force balance
\begin{equation}
\fbox{ \parbox{2cm} {\[
f=0.
  \]}}
\label{vpower26b}
\end{equation}
The requirement that  $f=0$ implies that  a variation  of $\lambda_b$ expends no internal power,   and  at first blush it appears that the  ``microforce balance'' \eqref{vpower26b}   is devoid of physical content. However, it does have    physical content, which is revealed later when we consider our thermodynamically consistent constitutive theory in Section~\ref{consttheory}.  As we shall see \eqref{vpower26b} will imply an internal constraint equation between $\lambda_b$ and the right Cauchy-Green tensor $\C$ and other constitutive variables  of the form $f(\C,\lambda_b,\dee,\nabla\dee)=0$, 
which will serve as an implicit equation for determining  $\lambda_b$ in terms of the  right Cauchy-Green tensor $\C$ and the other constitutive variables; cf. Section~\ref{intconstraint}.

\item 

Next, consider a generalized virtual velocity with
$\tilde{\y}={\bf 0}$ and   $\lambdabtilde=0$, and choose the virtual field $\deetilde$
arbitrarily.
The power balance \eqref{vpower4}  then yields
the   microscopic virtual-power relation
\begin{equation}
 \int\limits_{\partial{}\p}\!\zeta(\bfn_\mat)\deetilde\,da_\mat 
=
 \int\limits_{\p}\Bigl(
\varpi\deetilde 
+\bfxi \cdot \nabla \deetilde
\Bigr)\, \,dv_\mat ,
\label{vpower26}
\end{equation}
to be satisfied for all  $\deetilde$ and all $\p$. 
Equivalently,
using the divergence theorem,
$$
 \int\limits_{\partial{}\p}\!\bigl(\zeta (\bfn)
- \bfxi \!\cdot\!\bfn_\mat\bigr)\deetilde\,da_\mat+\int\limits_{\p}\bigl(
      \Div \,\bfxi -\varpi\bigr)\deetilde\,dv_\mat=0,
$$
and a standard argument yields the  {microscopic traction
condition}
\begin{equation}
\fbox{ \parbox{3cm} {\[  
\zeta(\bfn_\mat)=\bfxi\!\cdot\!\bfn_\mat,
\]}}   \label{vpower27}
\end{equation}
and the  {microscopic force balance}
\begin{equation}
\fbox{ \parbox{3.5cm} {\[
  \Div \,\bfxi -\varpi =0.
\]}}   \label{vpower28}
\end{equation}
\end{enumerate}

%

%


Finally, using the traction conditions  \eqref{vpower12} and \eqref{vpower26}, the actual   external expenditure  of power is
\begin{equation}
\calW_{\text{ext}}(\p)   =
\int\limits_{\partial\p}\!(\T_\mat\bfn_\mat) \cdot\dot\y \,da_\mat +
\int\limits_{\p}\!\bfb_\mat\cdot\dot\y\,dv_\mat
  +
 \int\limits_{\partial\p}\!(\bfxi\cdot\bfn_\mat)\vs\deedot\,da_\mat.
  \label{vpower29}
\end{equation}
%


As is standard, the Piola stress $\T_\mat$ is related to  the   symmetric Cauchy stress   $\bfT$  in the deformed body  by
\begin{equation}
  \bfT_\mat = J\, \T  \,\bfF^{-\trans},
  \label{cauchy1}
\end{equation}
so that
\begin{equation}
  \T=J^{-1}\T_\mat\F^{\trans}.
  \label{cauchy2}
\end{equation}
Further, it is convenient to  introduce  a new stress measure  called the  second Piola stress
\begin{equation}
\fbox{ \parbox{5cm} {\[
\T_{\mat\mat} \Def  \F^\inv \T_\mat =J\F^\inv \T \F^\invtrans, 
  \mskip3mu \]}} \label{2pk}
\end{equation}
which is symmetric. 


%
Next, differentiating   \eqref{kin10}$_1$ results in the following
expression for the rate of change of $\C$,
$$
\Cdot 
 =   \FT\Fdot
 +\Fdot{}^{\trans} \F.
$$
Hence, since $\T_{\mat\mat}$ is symmetric,
$$
\T_{\mat\mat}  \tendot\Cdot  = 2\T_{\mat\mat}\tendot  \FT\Fdot
=  2(\F \T_{\mat\mat})\tendot   \Fdot, 
$$
and upon using \eqref{2pk},
the stress power $\T_\mat\tendot\Fdot$ may be alternatively written as 
\begin{equation}
\T_\mat\tendot\Fdot= \onehalf  \T_{\mat\mat} \tendot \Cdot.
\label{vpower29a}
\end{equation}
Thus  the corresponding  actual internal expenditure of power \eqref{vpower2} may be written as 
%
%
\begin{equation}
\calW_{\text{int}}(\p)   =
\int\limits_{\p}\!\Big(   \onehalf\T_{\mat\mat} \tendot \Cdot   
+f\lambdabdot +\varpi\deedot
+\bfxi\!\cdot\nabla\deedot
\Big)\,dv_\mat.
 \label{vpower30}
\end{equation}
%


\subsection{Free-energy imbalance}
\label{energy0}

We  develop 
the theory
within a framework that accounts for the first two laws of thermodynamics. For isothermal processes the
first two laws typically collapse into a single dissipation inequality which  asserts that temporal changes in
free energy  of a part $\p$  be not greater than the power expended on    $\p$  \citep[cf., e.g.,][]{gurtin-fried-anand}.
Thus, let $\psi_\mat(\X,t)$  denote the  free energy density per unit reference volume.
Then,   the    \emph{free-energy imbalance under isothermal conditions}  requires
that  for each part  $\p$ of $\B$,
\begin{equation}
\overset{\cdot}{\overline{\int\limits_{\p}  \psi_\mat   \,dv_\mat}}\,\, \le  \,
\calW_{\text{ext}}(\p). \label{fei1}
\end{equation}

  Bringing the time derivative in \eqref{fei1} inside the integral,  using  $\calW_{\text{ext}}(\p)=\calW_{\text{int}}(\p)$,  eq. \eqref{vpower30}, and rearranging gives
 \begin{equation}
 \int_{\p}\Big( \dot\psi_\mat  
-  \onehalf\T_{\mat\mat} \tendot \Cdot
-f\lambdabdot
  -\varpi\deedot
- \bfxi\!\cdot\nabla\deedot\Big)\,dv_\mat \le 0.
 \label{linteint1}
 \end{equation}
Thus, since $\p$ was arbitrarily chosen, we obtain the following  local form  of
the  free-energy imbalance,
\begin{equation}
\fbox{ \parbox{7cm} {\[
\dot\psi_\mat  
- \onehalf\T_{\mat\mat} \tendot \Cdot
-f\lambdabdot
  -\varpi\deedot
- \bfxi\!\cdot\nabla\deedot
   \le 0.
\mskip3mu \]}}
\label{diss1}
\end{equation}
For later use 
we define the   {dissipation density} $\calD\ge 0$  
per unit  reference volume per unit time by
\begin{equation}
\fbox{ \parbox{7cm} {\[
\calD= 
 \onehalf\T_{\mat\mat} \tendot \Cdot  +f\lambdabdot  +\varpi\deedot
+ \bfxi\!\cdot\nabla\deedot
 - \dot\psi_\mat \ge 0.
 \mskip3mu \]}}
\label{calD}
\end{equation}

\begin{Remark}

For brevity we have not discussed the transformation properties under a change in frame  of the various fields appearing in our theory. 
Here, we simply note that all quantities in the free energy imbalance \eqref{diss1}  are invariant under a change in frame
\citep{gurtin-fried-anand}.
\end{Remark}

\subsection{Constitutive theory}
\label{consttheory}


 %
 Let $\bfLambda$ represent  the list
  \begin{equation}
  \bfLambda=\{\C,\lambda_b,\dee,\nabla\dee\}.
  \label{list}
  \end{equation}
Guided by\eqref{diss1}, we beginning by assuming  constitutive equations 
  for the free energy $\psi_\mat$,    the stress $\T_{\mat\mat}$,  the  scalar microstress $f$ and  $\varpi$, and the vector microstress $\bfxi$  are given by the constitutive equations
 \begin{equation}
 \begin{split}
  \psi_\mat &= \hat{\psi}_\mat(\bfLambda),  \qquad
    \T_{\mat\mat}  =\hat{\bfT}_{\mat\mat}(\bfLambda),\qquad
    f=\hat{f}(\bfLambda),\qquad
   \varpi  = \hat{\varpi}(\bfLambda),\qquad \bfxi=\hat{\bfxi}(\bfLambda).
 \end{split}
  \label{cons0}
  \end{equation}
 Then, 
 \begin{equation}
 \dot\psi_\mat = 
  \pards{\hat{\psi}_\mat(\bfLambda)}{\C}\tendot\Cdot
  +%
  \pards{\hat{\psi}_\mat(\bfLambda)}{\lambda_b}\dot\lambda_b
 +  \pards{\hat{\psi}_\mat(\bfLambda)}{\dee} \deedot
 +  
  \pards{\hat{\psi}_\mat(\bfLambda)}{\nabla\dee} \cdot\nabla\deedot.
    \label{cons3aa}
  \end{equation}
Using \eqref{cons3aa} and substituting the constitutive equations \eqref{cons0}    into the
free-energy imbalance \eqref{diss1}, we find that it may then be written as
  \begin{equation}
  \left[
  \pards{\hat{\psi}_\mat(\bfLambda)}{\C}-  \onehalf \hat{\T}_{\mat\mat}(\bfLambda)\right]\tendot\Cdot
  +\left[
  \pards{\hat{\psi}_\mat(\bfLambda)}{\lambda_b}-  \hat{f}(\bfLambda)\right]\dot\lambda_b
 +  \left[\pards{\hat{\psi}_\mat(\bfLambda)}{\dee} -\varpi \right]\deedot
 +  \left[
  \pards{\hat{\psi}_\mat(\bfLambda)}{\nabla\dee}-   \bfxi \right]\cdot\nabla\deedot 
  \le 0.
  \label{cons3a}
  \end{equation}
We introduce  an  \emph{energetic macrostress} $(\T_{\mat\mat})_\en$, and \emph{energetic  microstresses}  $\fen$, $\varpien$,  and $\xien$ through
\begin{equation}
\begin{split}
(\T_{\mat\mat})_\en &\Def \pards{\hat{\psi}_\mat(\bfLambda)}{\C},\qquad
\fen    \Def \pards{\hat{\psi}_\mat(\bfLambda)}{\lambda_b},\qquad
\varpien  \Def \pards{\hat{\psi}_\mat(\bfLambda)}{\dee},\qquad
\xien   \Def \pards{\hat{\psi}_\mat(\bfLambda)}{\nabla\dee}, 
\end{split}
\label{cons5b}
\end{equation}
and guided by \eqref{cons3a}  also introduce a  \emph{dissipative microstress} 
$\fdiss $, $\varpidiss $, and $\xidiss $   through
\begin{equation}
\begin{split}
(\T_{\mat\mat})_\diss &\Def \T_{\mat\mat}- (\T_{\mat\mat})_\en,\qquad
\fdiss   \Def f-\fen, \qquad  
\varpidiss   \Def \varpi-\varpien, \qquad  
 \xidiss  \Def  \bfxi-\xien.
\end{split}
\label{cons5c}
\end{equation}
Using \eqref{cons5b} and \eqref{cons5c}, leads to  the following reduced dissipation inequality
  \begin{equation}
  (\T_{\mat\mat})_\diss \tendot\Cdot
   +\fdiss\lambdabdot
 +\varpidiss\deedot
+ \xidiss\cdot\nabla\deedot 
 \ge  0.
  \label{cons5}
  \end{equation}
%
%

%

Next,  as  (special)  constitutive equations for  $(\T_{\mat\mat})_\diss$ $\fdiss$, $\xidiss$, and  $\varpidiss$  we assume that  tensor macrostress $\T_{\mat\mat}$, the  scalar microstress $f$ and the vector microstress $\bfxi$ are \emph{purely energetic} so that
\begin{equation}
\begin{split}
 (\T_{\mat\mat})_\diss &=\zed, \qquad
\fdiss  =0,\qquad
\xidiss  = \zed,
\end{split}
\label{dissipeqsa}
\end{equation}
while $\varpidiss$ is given by
\begin{equation}
\begin{split}
\varpidiss &=\alpha + \zeta\, \deedot , \qquad \text{with}\qquad \alpha  = \hat{\alpha}(\bfLambda) >0 , \quad \text{and} \quad  \zeta   = \hat{\zeta}(\bfLambda) >0,\end{split}
\label{dissipeqs}
\end{equation}
%
%
 so that   the dissipation inequality \eqref{cons5}   is satisfied, that is
 \begin{equation}
\calD= \left(\alpha  + \zeta\, \deedot  \right)\deedot    > 0\quad \text{whenever}  \quad \deedot > 0.
\label{mechdissip2}
\end{equation}
%


From \eqref{cons5b}, \eqref{cons5c}, \eqref{dissipeqsa}, and \eqref{dissipeqs} the tensor macrostress  $\T_{\mat\mat}$,  the scalar microstress  $f$, $\varpi$, and the vector microstress $\bfxi$ are given by the  thermodynamically consistent constitutive equations 
\begin{equation}
\begin{split}
\T_{\mat\mat} & = \underbrace{\pards{\hat{\psi}_\mat(\bfLambda)}{\C}}_{\text{energetic}},\qquad
f =  \underbrace{\pards{\hat{\psi}_\mat(\bfLambda)}{\lambda_b}}_{\text{energetic}},\qquad
\varpi = \underbrace{\pards{\hat{\psi}_\mat(\bfLambda)}{\dee}}_{\text{energetic}}+ \underbrace{\hat\alpha(\bfLambda)+ \hat\zeta(\bfLambda) \deedot }_{\text{dissipative}},\qquad
\bfxi  =  \underbrace{\pards{\hat{\psi}_\mat(\bfLambda)}{\nabla\dee}}_{\text{energetic}}.
\end{split}
\label{genac1}
\end{equation}

\subsection{Implicit equation for the bond deformation stretch}
\label{intconstraint}
The microforce balance \eqref{vpower26b}, viz. 
\begin{equation}
f=0,
 \label{vpower26aa}
\end{equation}
 together with the constitutive equation  \eqref{genac1} $_1$ gives the thermodynamic constraint
 \begin{equation}
 \pards{\hat{\psi}_\mat(\bfLambda)}{\lambda_b}=0,
  \label{mao1}
\end{equation}
 which serves as an implicit equation for $\lambda_b$, in terms of the other constitutive variables $(\C,\dee,\nabla\dee)$.
 
 
\subsection{Evolution equation for the phase field}

The microforce balance \eqref{vpower28}, viz. 
\begin{equation}
  \Div \,\bfxi -\varpi =0,
 \label{vpower28aa}
\end{equation}
  together with the constitutive equations \eqref{genac1} gives the evolution equation for the phase-field  variable $\dee$ as
\begin{equation}
\begin{split}
  \zeta \deedot    & =   F     \qquad \text{for} \qquad \deedot >0, \qquad \text{where}\\
F &\Def  \left[-\pards{\hat{\psi}_\mat(\bfLambda)}{\dee}+ \Div\left(\pards{\hat{\psi}_\mat(\bfLambda)}{\nabla\dee}  \right)\right]  -\hat\alpha(\bfLambda).
\end{split}
\label{genac2}
\end{equation}
Since $\zeta$ is positive-valued,  $F$ must be  positive for $\deedot$ to be positive and   the damage to increase.


\begin{Remark}
To formulate  a rate-independent theory, instead of \eqref{dissipeqs},  as 
 a special  constitutive equation  for $\varpidiss$,   we assume that  
\begin{equation}
\begin{split}
\varpidiss &=\alpha  \qquad  \text{with} \qquad   \alpha  = \hat{\alpha}(\bfLambda) >0,  \end{split}
\label{dissipeqs2}
\end{equation}
and in this case  eq. \eqref{genac2} reduces to the requirement that
\begin{equation}
\begin{split}
F& =   0     \qquad \text{for} \qquad \deedot >0 ;
\end{split}
\label{genac2a}
\end{equation}
 that is $F=0$ is a necessary condition for $\deedot>0$. Thus, in the rate-independent limit  we have
\begin{equation}
\deedot \ge0, \qquad F\le 0, \qquad \deedot\,F=0,
\label{genac5a}
\end{equation}
which are the  Kuhn-Tucker conditions associated with damage evolution.
It may be shown that  in the rate-independent limit,  $\deedot>0$ if and only if  the \emph{consistency condition}
\begin{equation}
 \dot{F}=0  \qquad \text{when}\qquad F=0 
\label{genac5}
\end{equation}
 is satisfied. The consistency  condition may be used to determine the value of $\deedot$ when it is non-zero.
 \end{Remark}

The theory formulated in this Appendix is summarized in Section~\ref{summary} in the main body of the paper.


\section{Some details of the numerical solution procedure}
\label{appendix:numerical}

We have implemented our theory using a finite element method within the commercial finite element code Abaqus \cite{abaqus}, through its user-defined element interface UEL.
Quasi-static plane stress problems are considered. Linear approximants on triangular and quadrangular elements are used for both the phase field $d$ and the deformation map components $\chi_i$, $i = 1, 2$. The phase field is represented within Abaqus by treating it as the temperature field and specifying analysis steps of type \texttt{*COUPLED TEMPERATURE-DISPLACEMENT}. Abaqus thus applies the backward Euler method for time integration of the microforce balance governing the phase field evolution \eqref{summgenac}; see the Abaqus theory manual \cite{abaqus}. Time integration accuracy is controlled with the option \texttt{DELTMX}, which specifies the maximum allowable nodal ``temperature'' change (actually the phase field value in this case) between time increments. We have found that a limit of $0.02$ provides a reasonable compromise between accuracy and computational efficiency.

The operator split  method   \citep{miehe2010phaseFieldImplement} is used to solve the linear momentum balance \eqref{macfb2} and the microforce balance \eqref{summgenac} in a staggered fashion; the staggered solution procedure is invoked from the Abaqus input file by declaring \texttt{*SOLUTION TECHNIQUE, TYPE=SEPARATED} in each load step. Default convergence criteria and tolerances are otherwise used.

The plane stress condition is enforced using an algorithm similar to that described in  \cite{KlinkelPlaneStress}. 
An implementation of the 3-dimensional version of the hyperelastic constitutive law of \citet{Mao17TalaminiEML} is used.
The condition
\[
    (T_\mat)_{33} = 0
\]
is solved for the unknown out of plane stretch $F_{33}$ with a Newton-Raphson iteration scheme, built on top of the constitutive law evaluation routine called during the assembly of the nodal residual forces and tangent stiffness matrices. After solution, static condensation of the material tangent stiffness operator is performed to get the plane stress tangent operator; see \citet{KlinkelPlaneStress} for details.

Considering our original development of hyperelastic constitutive law for the Piola stress in \citep{Mao17TalaminiEML}, the major difference here is the appearance of the $g(d)$ term in the implicit equation for the bond stretch \eqref{summary_condition}. However, with the staggered update procedure, $d$ is considered as a constant in the stress constitutive law evaluation, and hence requires no additional Jacobian terms. 
As noted in footnote \ref{foot:regularization} above, in computations, we modify the degradation function to 
\[
    g(d) = (1-d)^2 + k,
\]
where $k$ is a small positive constant. This prevents complete loss of stress-bearing capacity of the material to avoid non-uniqueness of the solution. We have used a value  of $k \sim 10^{-4}$ in the simulations presented here.

For each call of the constitutive update at each quadrature point, we solve the implicit nonlinear equation \eqref{summary_condition} for $\lambda_b$ with Newton-Raphson iteration supplemented with bisection using the \texttt{rtsafe} routine of \citet{press1987numerical}.

\bibliography{elastomer_fracture}

\end{document}